\documentclass[journal]{IEEEtran}

\usepackage{graphicx, dblfloatfix}
\usepackage{amsmath}
\usepackage{amssymb}
\usepackage{graphics}
\usepackage{cite}
\usepackage{amsfonts}
\usepackage{textcomp}
\usepackage{multicol}
\usepackage{multirow}
\usepackage{color}
\usepackage{fixltx2e}
\usepackage{epstopdf}
\usepackage{bm}
\usepackage{siunitx}

\begin{document}
\bstctlcite{IEEEexample:BSTcontrol}

\title{All-Analog Adaptive Equalizer for Coherent Data Center Interconnects}

\author{Nandakumar Nambath, Rakesh Ashok, Sarath Manikandan, Nandish Bharat Thaker, Mehul Anghan,\\Rashmi Kamran, Saurabh Anmadwar, and Shalabh Gupta
\thanks{This work was supported by the Ministry of Electronics and Information Technology, Government of India. Grant No. R-23011/15/2013-CC\&BT.

N. Nambath is with the School of Electrical Sciences, Indian Institute of Technology Goa, Ponda -- 403401, India (e-mail: npnandakumar@iitgoa.ac.in).

R. Ashok, R. Kamran, and S. Gupta are with the Department of Electrical Engineering, Indian Institute of Technology Bombay, Mumbai -- 400076, India (email: rakesh.ashok@iitb.ac.in; rashmikamran@ee.iitb.ac.in; shalabh@ee.iitb.ac.in).

S. Manikandan is with Maxlinear Technologies Pvt. Ltd., Bengaluru -- 560103, India (email: smanikandan@maxlinear.com).

N. B. Thaker and S. Anmadwar are with Intel Technology India Pvt. Ltd., Bengaluru -- 560103, India (e-mail: nandish.bharat.thaker@intel.com; saurabh.r.anmadwar@intel.com).

M. Anghan is with Rambus Chip Technologies India Pvt. Ltd., Bengaluru -- 560029, India (e-mail: manghan@rambus.com).

S. Manikandan, N. B. Thaker, M. Anghan, and S. Anmadwar were with the Department of Electrical Engineering, Indian Institute of Technology Bombay, Mumbai -- 400076, India when this work was carried out.}}

\maketitle

\begin{abstract}

  In a high-speed coherent optical transmission system, typically the signals obtained at the receiver front-end are digitized using very high-speed ADCs and then processed in the digital domain to remove optical channel impairments. In this work, we show that these signals can instead be processed in the analog domain, which can significantly reduce the power consumption and the complexity of the receiver. The first all-analog adaptive equalizer for receivers of coherent dual-polarization optical links has been presented with its detailed architecture and measurement results. The proof-of-concept equalizer uses the constant modulus algorithm for blind adaptation of its weight coefficients to implement a 4$\times$4 2-tap FIR filter in 130\,nm SiGe BiCMOS technology. Its functionality is evaluated experimentally for 40\,Gb/s data rate and 10\,km standard single mode fiber channel. This demonstration shows that the use of all-analog processing for short reach data-center interconnects is feasible and is a much simpler solution than the use of the high-speed ADC+DSP based approach. Moreover, when implemented in advanced CMOS or FinFET technologies, the power consumption of the equalizer is expected to be significantly lower than the DSP based implementations in similar process technologies. 
  
\end{abstract}

\begin{IEEEkeywords}
  
  Analog coherent receiver, adaptive equalizer, analog signal processing, BiCMOS integrated circuits.
  
\end{IEEEkeywords}

\IEEEpeerreviewmaketitle

\section{Introduction}
  
  \IEEEPARstart{D}{ata} center interconnects (DCIs) are expected to carry around 80\% of the global Internet traffic, which is estimated to increase two-fold from today's traffic of 200\,EB/month, by 2022 \cite{Cisco_20162021, Cisco_20172022}. To support the increased traffic, DCI interface speeds are projected to cross 1.6\,Tb/s in the near future \cite{Ethernet_2018}. To meet the current needs, 4-level pulse amplitude modulation (PAM-4) format based intensity modulation-direct detection (IMDD) transceivers with electronics operating at 56\,GBd are being standardized \cite{FM_OFC2017,Ethernet_2018}. However, IMDD techniques are anticipated to give way to the transmission of more spectrally efficient modulation formats through single-mode fiber (SMF) in short-reach links \cite{EP2017, SC2019}, mainly due to data rate scalability limitations \cite{SM2015}. As opposed to IMDD, coherent techniques aid in achieving data rate scalability along with a better receiver sensitivity \cite{KK2016}. Coherent links that use higher order modulation formats and polarization multiplexing help to achieve higher capacities and longer channel lengths, relying heavily on digital signal processing (DSP) to overcome the impairments added by the fiber channel. However, such links require high-speed, high-precision analog-to-digital converters (ADCs) to digitize the received signals after optical-to-electrical (O/E) conversion followed by high-speed DSP. Therefore, usage of the ADC+DSP approach becomes prohibitive for DCIs due to a huge amount of power consumption and complexity involved in this solution \cite{IE2018}.
  
  To overcome this issue an analog domain processing based receiver was first proposed in our previous work \cite{PM2012}. Schematic level circuit simulations results of the receiver were presented in \cite{NN2015} and the low-speed measurement results of an equalizer chip for the receiver were presented in \cite{NN2017}. Since then a few other solutions using analog signal processing (ASP) for coherent optical links were reported\cite{PJ2017, MM2018, IE2018, CS2019, LM2013, VJ_JLT2018}. For example, \cite{PJ2017} discusses an analytical study on the feasibility of DSP-free homodyne dual-polarization quadrature phase-shift keying (DP-QPSK) receivers for DCIs. The reported architecture uses a simplified $2\times2$ equalizer, to correct small amounts of dispersion and bandwidth limitations, the weight coefficients of which can be updated by using either the constant modulus algorithm (CMA) or a least mean square algorithm. In another study, validation results of a proof-of-concept DSP-free coherent transceiver for DCIs that uses dual polarization-16 quadrature amplitude modulation (DP-16QAM) format was reported \cite{MM2018}. Recently, an analog coherent engine, working up to 400\,Gb/s dissipating 2\,W of power, that uses ASP followed by a feed-forward equalizer filter to improve signal quality for 15\,m SMF was demonstrated \cite{IE2018}.
  
  Studies on low-power analog coherent optics, based on optical phase-locked loop (OPLL), which work for very short distances due to the choice of low dispersion, but high attenuation O-band were also reported \cite{CS2019, SC2019}. Another work demonstrates a 40\,Gb/s OPLL based analog coherent binary phase-shift keying receiver that uses the integration of electronic and photonic integrated circuits (ICs) to eliminate the usage of DSP for carrier offset removal \cite{LM2013}. Recently, a technique to achieve carrier phase synchronization using ASP based IC, phase modulator, and tunable lasers was demonstrated \cite{RA_OFC2019, RA2020}. Receivers in \cite{LM2013, RA_OFC2019} use OPLL for phase and frequency error correction to achieve significant power savings by removing DSP completely.
    
  \begin{figure}[t!]
    \centering
    \includegraphics[scale=0.7]{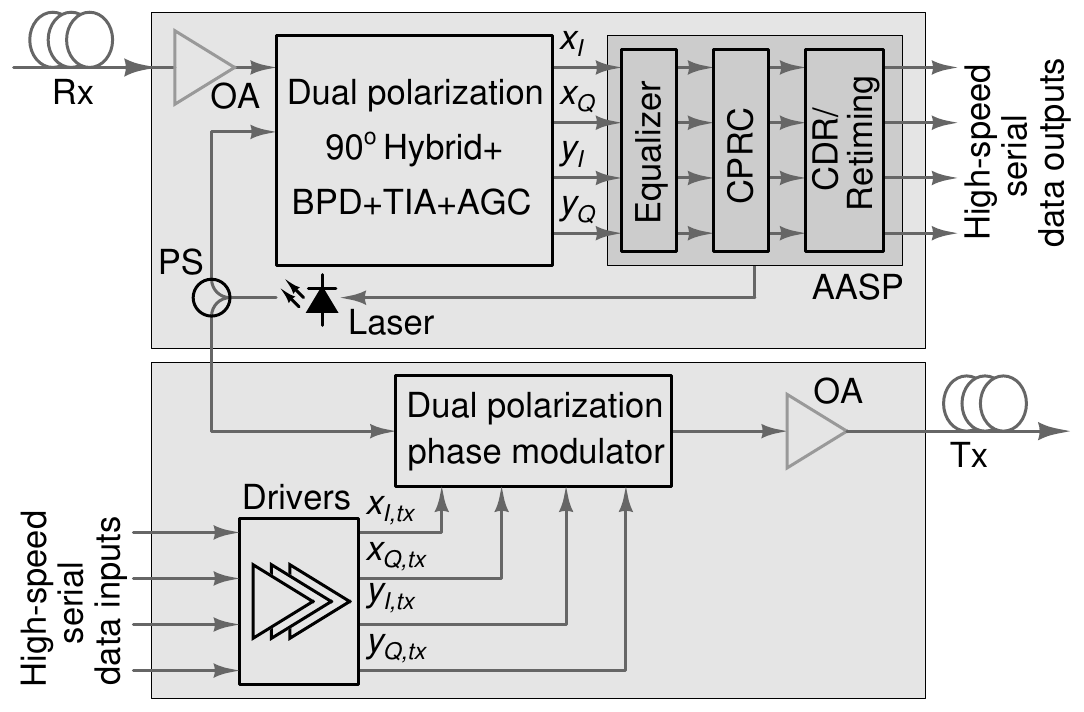} 
    \caption{Block diagram of the proposed transceiver for coherent optical links with a dual polarization, phase modulated carrier. BPD: balanced photo-diode, TIA: trans-impedance amplifier, AGC: automatic gain control, PS: power splitter, AASP: all-analog signal processing, CPRC: carrier phase recovery and compensation, CDR: clock and data recovery, and OA: optical amplifier.}
    \label{fig:transceiver}   
  \end{figure}
    
  From  the literature, it can be summarized that miniaturized analog coherent transceiver using electronic and photonic integration is a solution for the power dissipation-size-cost problem in the future DCIs. The block diagram of such a transceiver is given in Fig. \ref{fig:transceiver}. In a coherent transceiver, a major portion of the power is consumed by the receiver electronics which can be reduced if the main signal processing operations--equalization, carrier phase recovery and compensation (CPRC), and clock and data recovery (CDR) are carried out using ASP as shown in the figure. In the receiver side, an equalizer is used to mitigate the effects of channel dispersion. A CPRC is used to compensate for the phase and frequency mismatches between the transmitter and receiver lasers, and a CDR is used for clock recovery from the received signals.
  
  Equalizer being the major power dissipating block in coherent optical receivers \cite{PB2014}, we implemented an all-analog equalizer for a proof-of-concept validation of the ASP based transceiver. It can also be seen from the literature that analog domain processing is an attractive choice for low-power equalization in various types of high-speed links \cite{HW2003, SP2007, JH2010, WK2013, EM2014, PQ2015, AS2015, PC2015, LD_JSSC2010}. Specifically for optical links, a CMOS receiver with a continuous-time linear equalizer for 30\,Gb/s links was reported in \cite{PQ2015} and a monolithic optoelectronic IC designed in a 130\,nm CMOS process that uses analog domain slope detection based adaptive equalizer was shown in \cite{LD_JSSC2010} for links with a carrier of 850\,nm wavelength. As a major milestone in this area, the first all-analog adaptive equalizer for DP-QPSK links was demonstrated in our previous work \cite{NN2017}. In this present work, we report the proof-of-concept demonstration of the equalizer operating at a rate as high as 40\,Gb/s. We also present the detailed architecture of the equalizer and discuss the practical constraints and future directions of the ASP approach.  

  \begin{figure*}[t!]
    \centering
    \begin{tabular}{l}
      \includegraphics[scale=0.7]{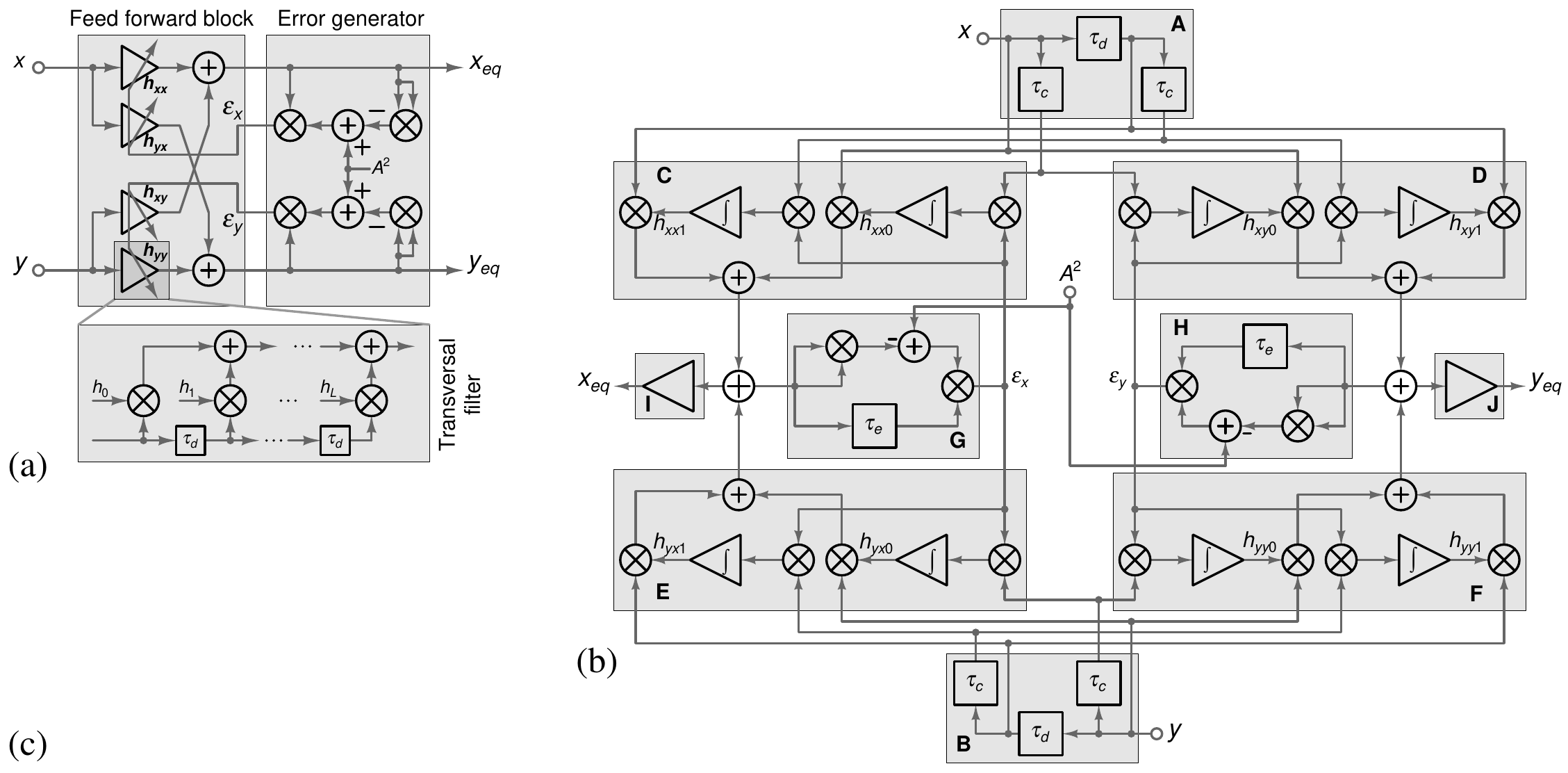}\\[-3.05cm] \hspace{1.2cm}\includegraphics[scale=0.21]{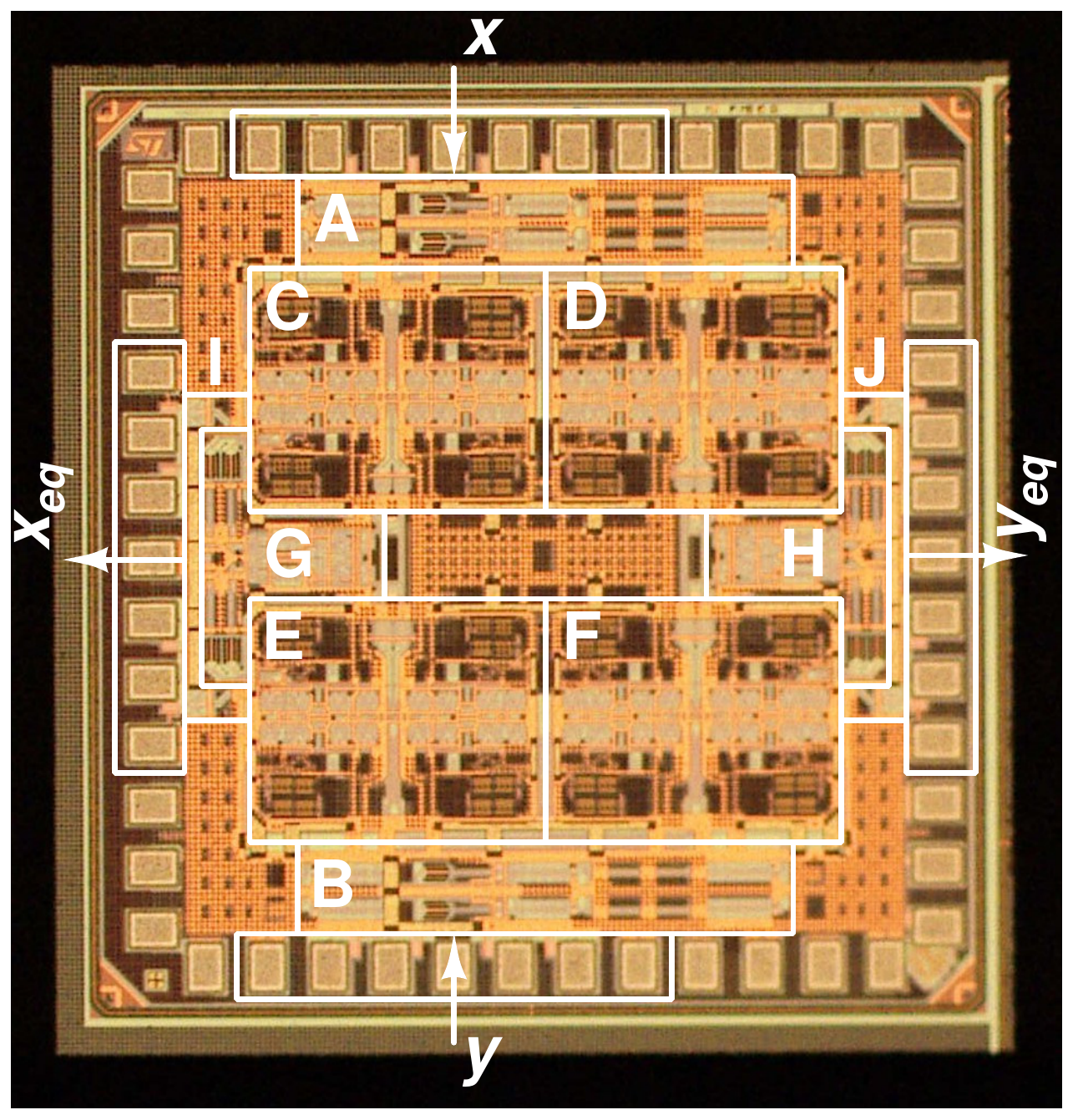}
    \end{tabular}
    \caption{CMA equalizer: (a) Architecture with the structure of a transversal filter on the bottom side. (b) Architecture of a two-tap equalizer. (c) IC micrograph. Blocks--A: X delay line, B: Y delay line, C: X input to X output forward path, D: X input to Y output forward path, E: Y input to X output forward path, F: Y input to Y output forward path, G: X error generator, H: Y error generator, I: X output buffer, and J: Y output buffer. $\tau_d$ is the tap delay, $\tau_c$ is the delay inserted to compensate for the delay mismatch in the forward paths, and $\tau_e$ is the delay inserted to compensate for the delay mismatch in the error generators.}
    \label{fig:cma_eq}   
  \end{figure*}
     
\section{System Overview}
\label{sec:overview}

  The optical-to-electrical (O/E) front-end of a DP-QPSK receiver gives out four electrical signals corresponding to the in-phase (I) and quadrature-phase (Q) components of the X and Y polarizations. These signals will not be independent of each other due to dispersion, polarization rotation, and polarization-dependent loss. Hence, a $4\times4$ multi-dimensional equalizer needs to be used to process all the signals jointly. The equalizer should adaptively mitigate the dispersion effects of the fiber, which is a time-varying phenomenon. To adapt the equalizer weight coefficients, the CMA algorithm \cite{SY1975} is used, which is one of the simplest blind equalization algorithms that can be used with DP-QPSK signals.
  
  Fig. \ref{fig:cma_eq}(a) shows the architecture of the equalizer that uses the CMA algorithm to adapt the weight coefficients. The equalizer has a feed-forward block and an error generator. It has inputs $x$ and $y$ and outputs $x_{eq}$ and $y_{eq}$ which are complex signals corresponding to the X and Y polarizations, respectively. The feed-forward block has four transversal filters with coefficients $\bm h_{xx}$, $\bm h_{xy}$, $\bm h_{yx}$, and $\bm h_{yy}$ which are arranged as a butterfly structure. This structure generates equalized output signals, $x_{eq}$ and $y_{eq}$ from the inputs, and the CMA error signals $\varepsilon_x$ and $\varepsilon_y$, which are calculated in the error generator. The outputs of the equalizer are given by \cite{NN2015} 
  \begin{equation}
    x_{eq} = \bm h_{xx}^T\bm x+\bm h_{xy}^T\bm y, y_{eq} = \bm h_{yx}^T\bm x+\bm h_{yy}^T\bm y,
  \label{eqn:sys}
  \end{equation}
  where $\bm x$ and $\bm y$ are vectors containing delayed input signals. The update equations of the equalizer weight are given by \cite{NN2015} 
  \begin{equation}
    h_{xx,k}(t) = \beta \int_0^{t} x_{eq}(\tau) [1-|x_{eq}(\tau)|^2 ]x(\tau-k\tau_d) d\tau,
  \label{eqn:contHxx}
  \end{equation}
  \begin{equation}
    h_{xy,k}(t) = \beta \int_0^{t} y_{eq}(\tau) [1-|y_{eq}(\tau)|^2 ]x(\tau-k\tau_d) d\tau,
  \label{eqn:contHxy}
  \end{equation}
  \begin{equation}
    h_{yx,k}(t) = \beta \int_0^{t} x_{eq}(\tau) [1-|x_{eq}(\tau)|^2 ]y(\tau-k\tau_d) d\tau,
  \label{eqn:contHyx}
  \end{equation}
  \begin{equation}
    h_{yy,k}(t) = \beta \int_0^{t} y_{eq}(\tau) [1-|y_{eq}(\tau)|^2 ]y(\tau-k\tau_d) d\tau,
  \label{eqn:contHyy}
  \end{equation}  
  \noindent where $\beta$ is a gain constant, $x=x_I+jx_Q$ and $y=y_I+jy_Q$ are the X and Y polarization input signals of the equalizer, $x_{eq}=x_{eq,I}+jx_{eq,Q}$ and $y_{eq}=y_{eq,I}+jy_{eq,Q}$ are the equalizer outputs in the corresponding polarizations, $\tau_d$ is the tap delay, and $0 \leqslant k \leqslant L$, where $L$ is the total number of delay cells in each transversal filter. The CMA algorithm is known to converge for any positive value of the gain constant $\beta$ \cite{MW1995}.
  
\section{Implementation Details}
\label{sec:impln}

  A fractionally spaced two-tap continuous-time CMA equalizer for 100\,Gb/s DP-QPSK links is designed and fabricated using 130\,nm SiGe BiCMOS technology from ST Microelectronics as a proof-of-concept. Fig. \ref{fig:cma_eq}(b) shows a detailed block diagram of the prototype equalizer, which consists of linear transversal filters, error generators, and weight-update modules. The figure is a block-level translation of the system described by (\ref{eqn:sys})-(\ref{eqn:contHyy}) with minor modifications to take care of the circuit level issues discussed in the following sub-section.
  
\subsection{Issues of Delay Mismatch and Signal Swing}  
\label{subsec:issues}

  Path delay is an inherent problem associated with any circuit which gets worse as the number of circuit blocks in the path is increased. This problem becomes crucial if the circuit has multiple parallel paths. Signal delays through the parallel paths are to be made equal over the desired frequency range. It is evident from (\ref{eqn:contHxx})-(\ref{eqn:contHyy}) that there are parallel paths in the equalizer. Signals through the parallel paths get multiplied down the signal path. In such cases, delay cells are inserted in the paths that have lower group delays. In Fig. \ref{fig:cma_eq}(b) delay cells $\tau_c$ are inserted to cancel the delay mismatch in the forward paths and $\tau_e$ are inserted to cancel the delay mismatch in the error generators. Equations (\ref{eqn:contHxx})-(\ref{eqn:contHyy}) assume a normalized single ended signal magnitude which is taken as 100\,mV in the circuit. Gains of all the building blocks are scaled according to this signal level. In Fig. \ref{fig:cma_eq}(b) the expected amplitude is represented by $A$, which is 100\,mV for a single-ended signal. Taking these into consideration, (\ref{eqn:contHxx}) can be modified as $h_{xx,k}(t) = \beta \int_0^{t} x_{eq}(\tau-\tau_e) [A^2-|x_{eq}(\tau)|^2 ]x(\tau-k\tau_d-\tau_c) d\tau.$
  Equations (\ref{eqn:contHxy})-(\ref{eqn:contHyy}) can also be modified in a similar way. It can be concluded from Fig. \ref{fig:cma_eq}(b) that the main operations in the equalizer are delay, multiplication, addition, and integration. The following sub-section briefly describes design details of basic building blocks of the equalizer.

\subsection{Basic Building Blocks of the Equalizer}

  Building blocks of the equalizer are designed, integrated, and simulated in Cadence Virtuoso analog design environment. All integrated circuits, be it DSP or ASP based, are prone to device mismatches, offset issues etc. These issues are taken care of during the design phase by doing Monte Carlo analysis which is a statistical technique used in circuit simulation to account for tolerance issues in the circuit components using Gaussian random numbers. All these analyses are performed at the block level design, various levels of integration, and the overall equalizer. Responses of the building blocks presented in this paper are results of post-layout Monte Carlo simulations with 200 runs at 27\textdegree C with 50\,fF parasitic capacitance at each output node. The simulations also include mismatches among circuit components, which is an important parameter when the system has differential circuits, and effects of process variations on device parameters such as the small signal current gain of BJTs and the threshold voltage of MOSFETs. Since the equalizer is adaptive, not all variations may be critical. These issues can also be handled well by using calibration techniques similar to the ones used in ADCs.
    
  \begin{figure}[t!]
    \centering
    \begin{tabular}{cc}
      (a) & \hspace{-1cm} \includegraphics[scale=0.35]{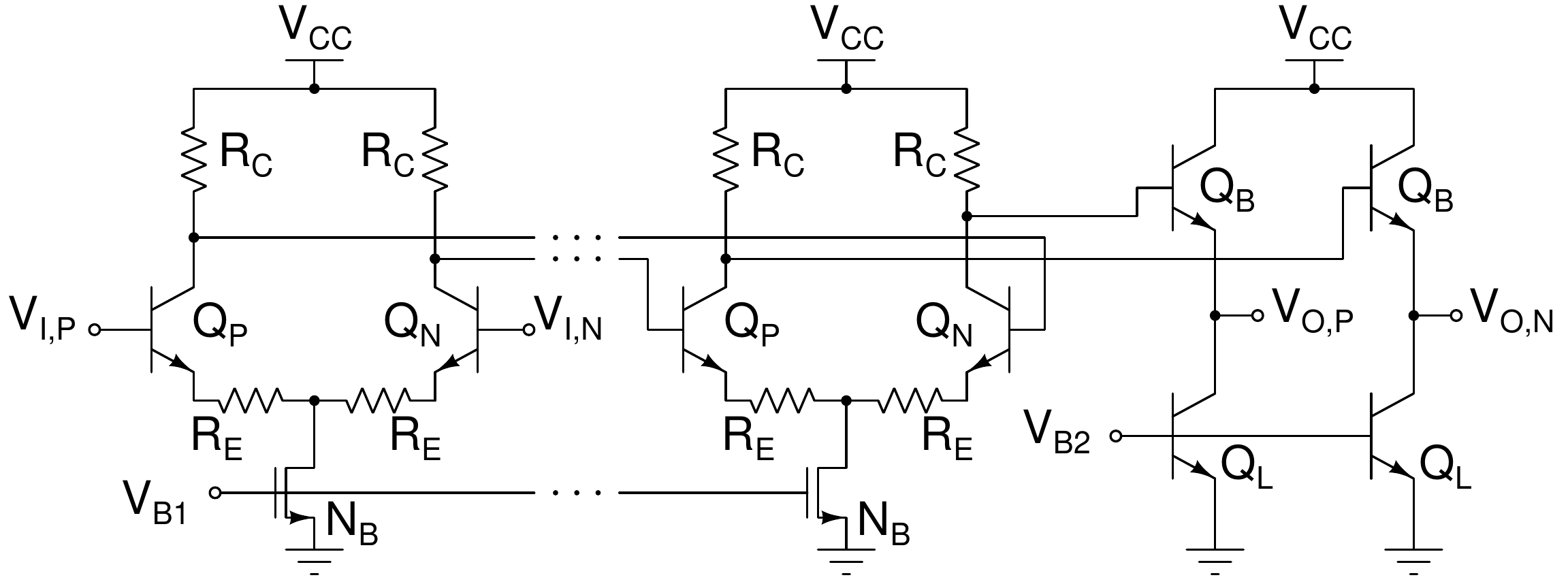}\\
      (b) & \hspace{-1cm} \includegraphics[scale=0.45]{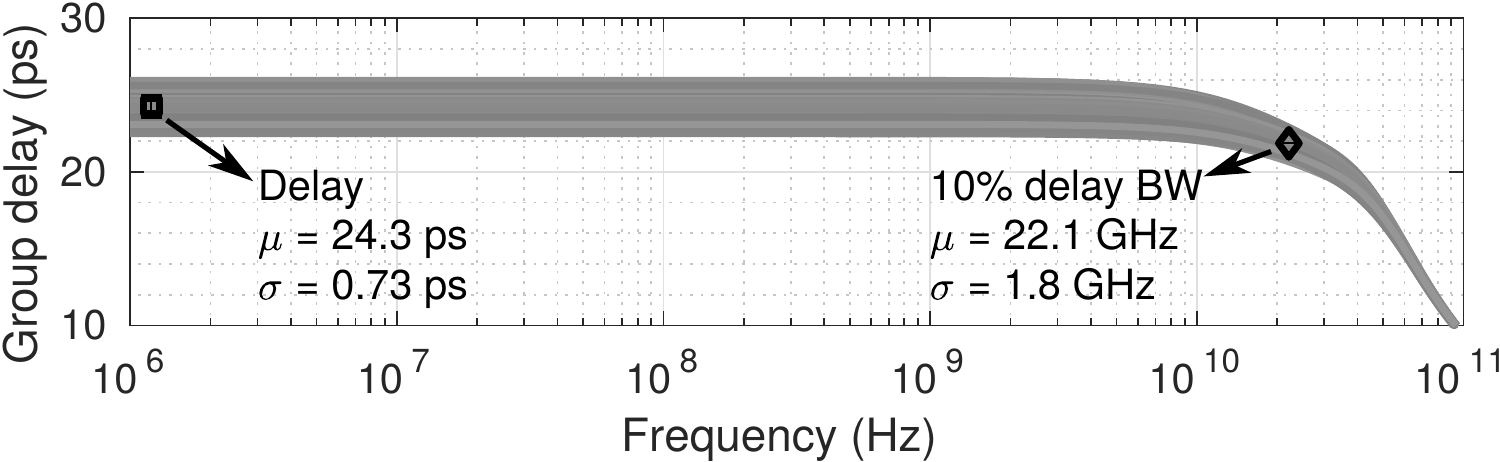}
    \end{tabular}    
    \caption{Delay cell: (a) Schematic. ${E_{L,Q_P,Q_N}}$ = 2\,\si{\micro\meter}, ${E_{L,Q_B}}$ = 1.6\,\si{\micro\meter}, ${E_{L,Q_L}}$ = 2.75\,\si{\micro\meter}, ${(W/L)_{N_B}}$ = 47/0.28, ${R_C}$ = 80\,$\Omega$, ${R_E}$ = 50\,$\Omega$, ${V_{B1}}$ = 0.8\,V, ${V_{B2}}$ = 0.9\,V, ${I_{N_B}}$ = 3\,mA, and ${I_{Q_L}}$ = 2.5\,mA. (b) Group delay response of the delay cell $\tau_d$.}
    \label{fig:delay_sch}
  \end{figure}  

\subsubsection{Delay Cell}

  An active delay cell is chosen over a passive delay cell to save chip area and hence, for the ease of routing. The delay cell is implemented by cascading several degenerated common emitter (CE) stages followed by a common collector (CC) buffer to drive a large capacitive load. A circuit schematic of the delay cell is shown in Fig. \ref{fig:delay_sch}(a). The group delay of the cell $\tau_d$ with five cascaded CE stages is shown in Fig. \ref{fig:delay_sch}(b). The delay cell, with an area of $\sim$95\,\si{\micro\meter}$\times$50\,\si{\micro\meter}, has a DC gain of \textminus1.4\,dB, and a bandwidth of 20.7\,GHz. This cell provides a group delay of 24.3\,ps with a 10\% delay-bandwidth of 22.1\,GHz. The delay cells $\tau_c$ and $\tau_e$ are implemented similarly by varying the number of CE stages as per the delay needed. The values of $\tau_c$ and $\tau_e$ are determined by doing AC analysis based group delay measurement and transient analysis in the Cadence Virtuoso analog design environment.  
    
  \begin{figure}[t!]
    \centering
    \begin{tabular}{ll}
      \includegraphics[scale=0.35]{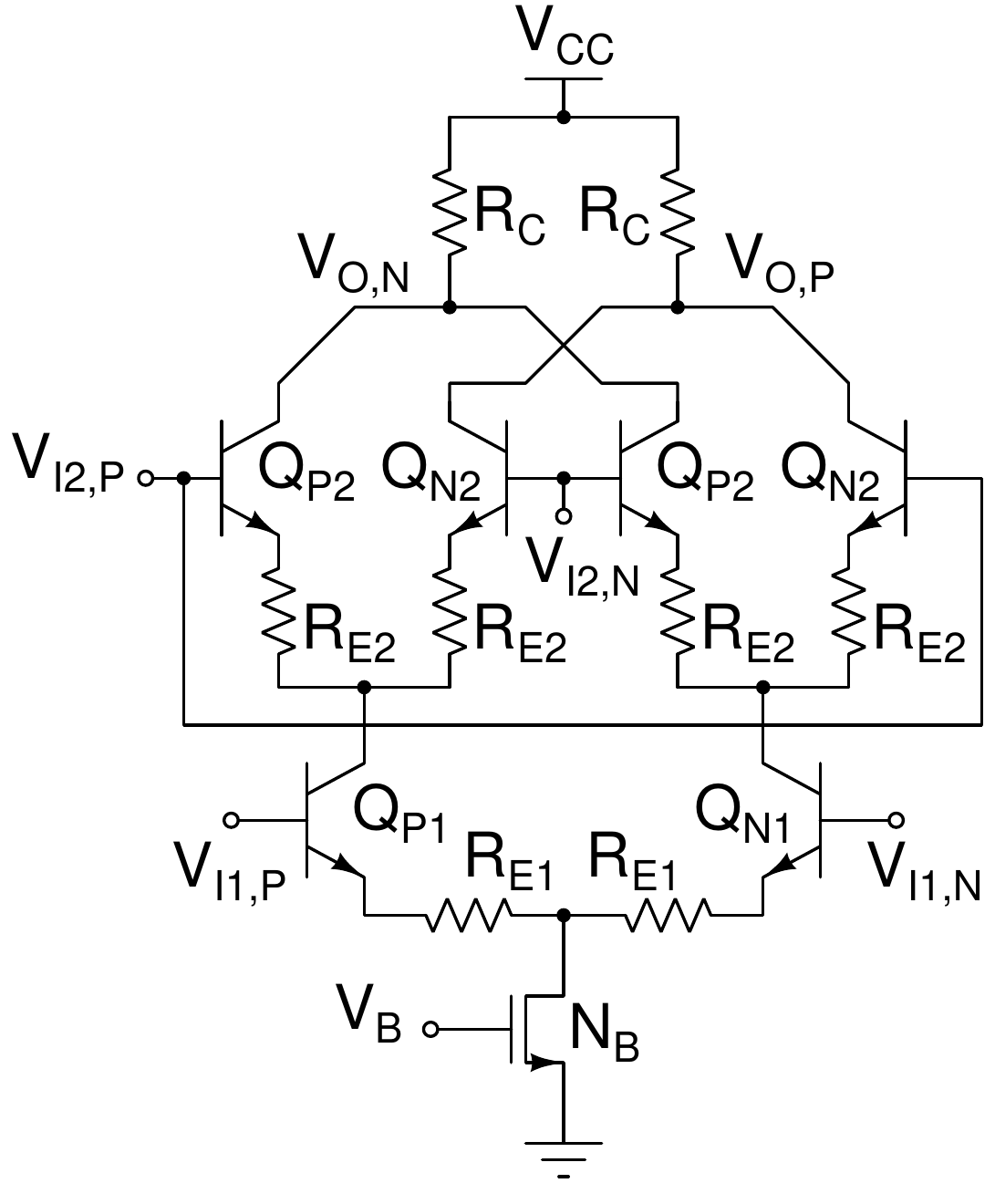} & \includegraphics[scale=0.45]{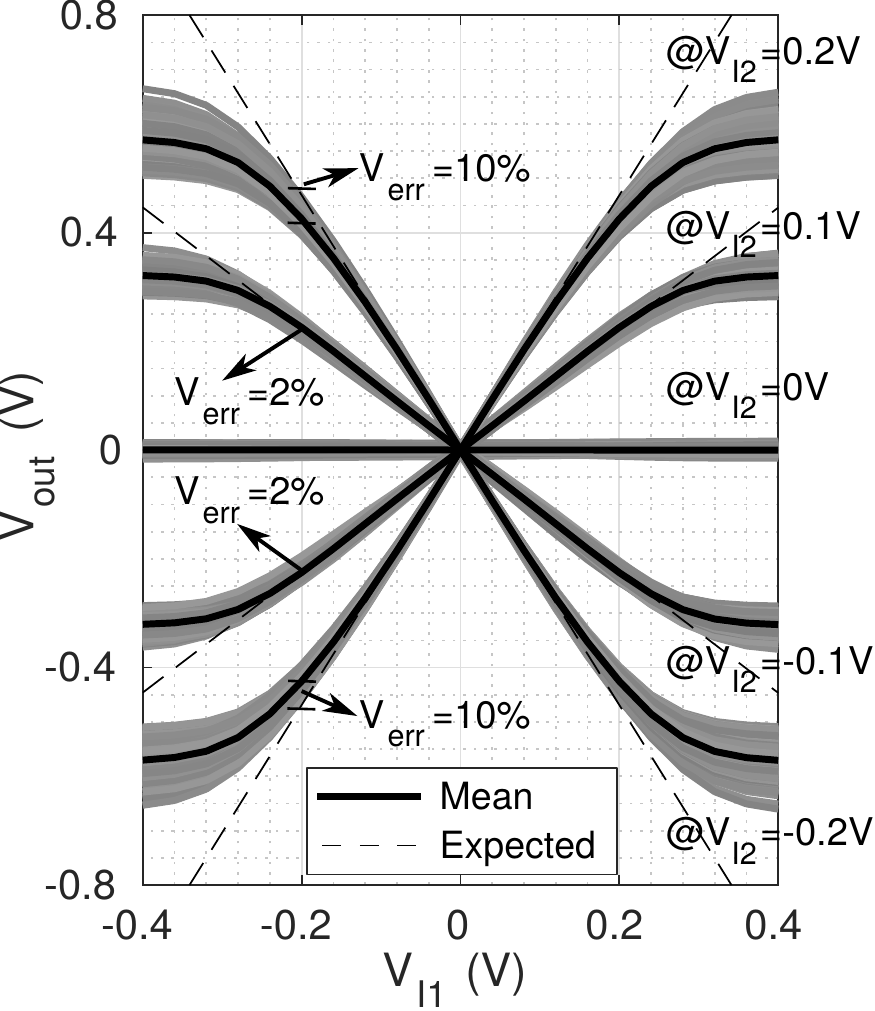}\\[-0.5cm]
      (a) & (b)
    \end{tabular}    
    \caption{Gilbert cell multiplier: (a) Schematic. ${E_{L,Q_{P1}}}$ = 6\,\si{\micro\meter}, ${E_{L,Q_{N1}}}$ = 6\,\si{\micro\meter}, ${E_{L,Q_{P2}}}$ = 3\,\si{\micro\meter}, ${E_{L,Q_{N2}}}$ = 3\,\si{\micro\meter}, ${(W/L)}_{N_B}$ = 170/0.28, ${R_C}$ = 100\,$\Omega$, ${R_{E1}}$ = 25\,$\Omega$, ${R_{E2}}$ = 10\,$\Omega$, ${V_B}$ = 0.66\,V, and ${I_{N_B}}$ = 6\,mA. (b) DC transfer characteristics of the multiplier.}
    \label{fig:mult_sch}
  \end{figure}

\subsubsection{Multiplier} 

  The multiplication operation is realized using the Gilbert cell topology \cite{R2002}. Fig. \ref{fig:mult_sch}(a) shows the schematic of the multiplier. The circuit has degeneration resistors for both the inputs to improve the linearity. The designed multiplier has a DC gain of 7.4\,dB and a bandwidth of 20.2\,GHz from I1 when I2 is 200\,mV, and a DC gain of 6.02\,dB and a bandwidth of 17.9\,GHz from I2 when I1 is 200\,mV. Due to the lower bandwidth of I2 low-speed signals such as filter weight coefficients are applied to this input. DC transfer characteristics of the multiplier is shown in Fig. \ref{fig:mult_sch}(b). It can be observed that the maximum deviation from the linear region is only 10\% in the signal swing of concern. A single multiplier occupies an area of $\sim$40\,\si{\micro\meter}$\times$45\,\si{\micro\meter}. A complex multiplier is implemented by connecting four such Gilbert cells, and a squaring circuit by giving the same signals to both the inputs of a Gilbert cell.    
  
  \begin{figure}[t!]
    \centering
    \includegraphics[scale=0.35]{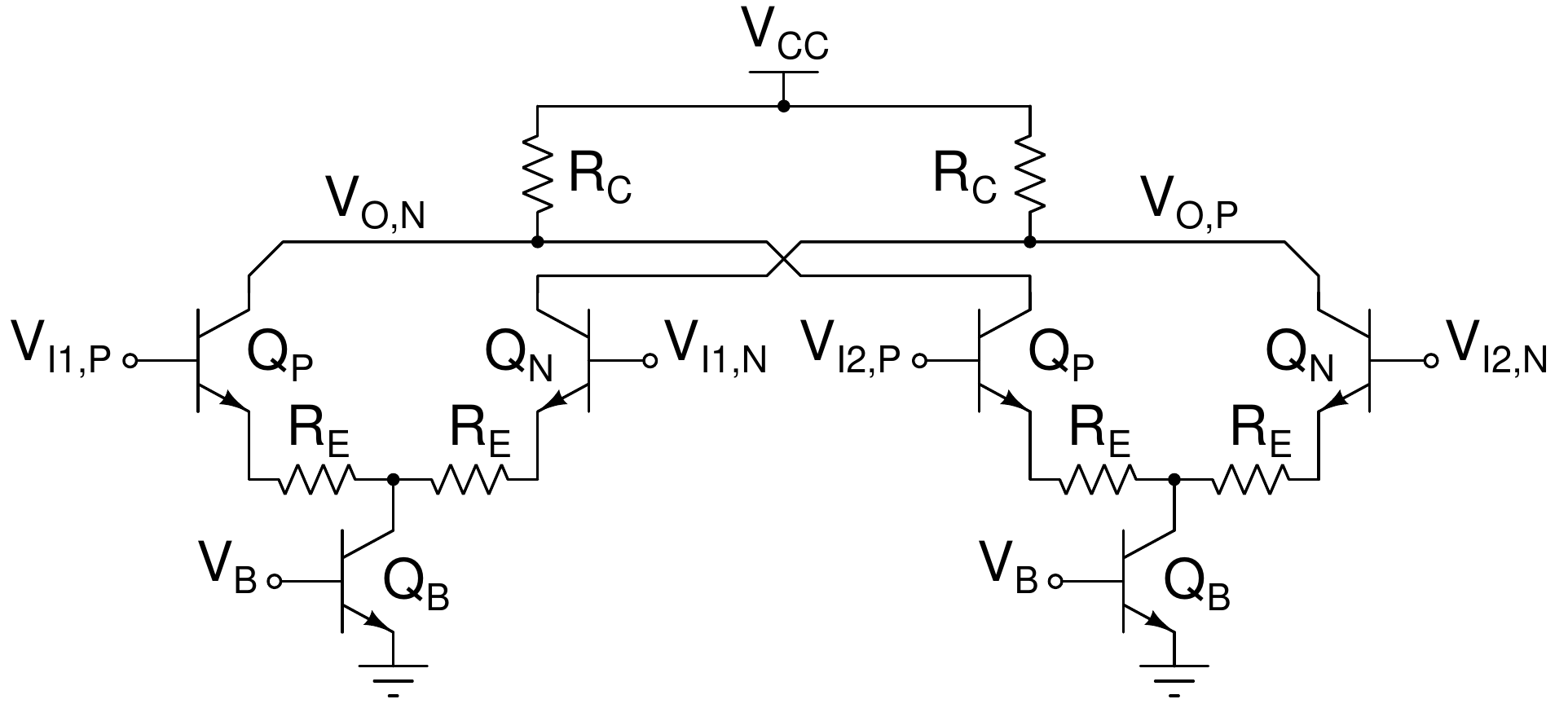}
    \caption{Schematic of the adder. ${E_{L,Q_P,Q_N, Q_B}}$ = 7.5\,\si{\micro\meter}, ${R_C}$ = 50\,$\Omega$, ${R_E}$ = 10\,$\Omega$, ${V_B}$ = 0.86\,V, and ${I_{Q_B}}$ = 6\,mA.}
    \label{fig:adder_sch}
  \end{figure}

\subsubsection{Adder}
   
  The addition operation is performed in the current domain by adding two current signals onto a common resistor. This technique is used in blocks such as complex multiplier where routing is minimal. When an additional gain is required, two degenerated CE sections are used with common load resistors, as shown in Fig. \ref{fig:adder_sch}. This adder has 12.3\,dB DC gain and 18.4\,GHz bandwidth. For the convenient placement in the equalizer layout, the adder is designed as half circuits, each of which has an area of $\sim$20\,\si{\micro\meter}$\times$40\,\si{\micro\meter}.
      
  \begin{figure}[t!]
    \centering
    \begin{tabular}{ll}
      \includegraphics[scale=0.35]{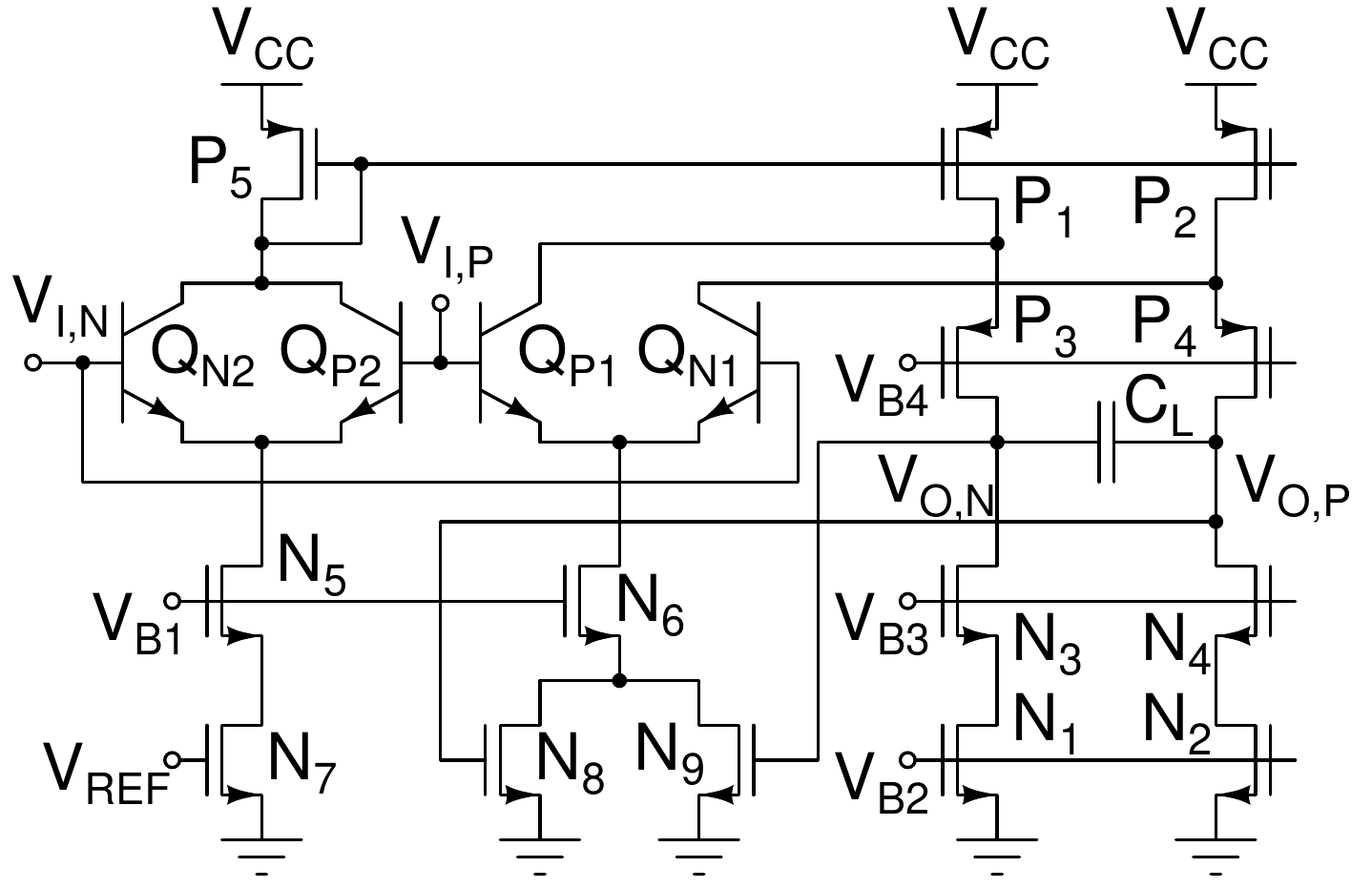} & \includegraphics[scale=0.35]{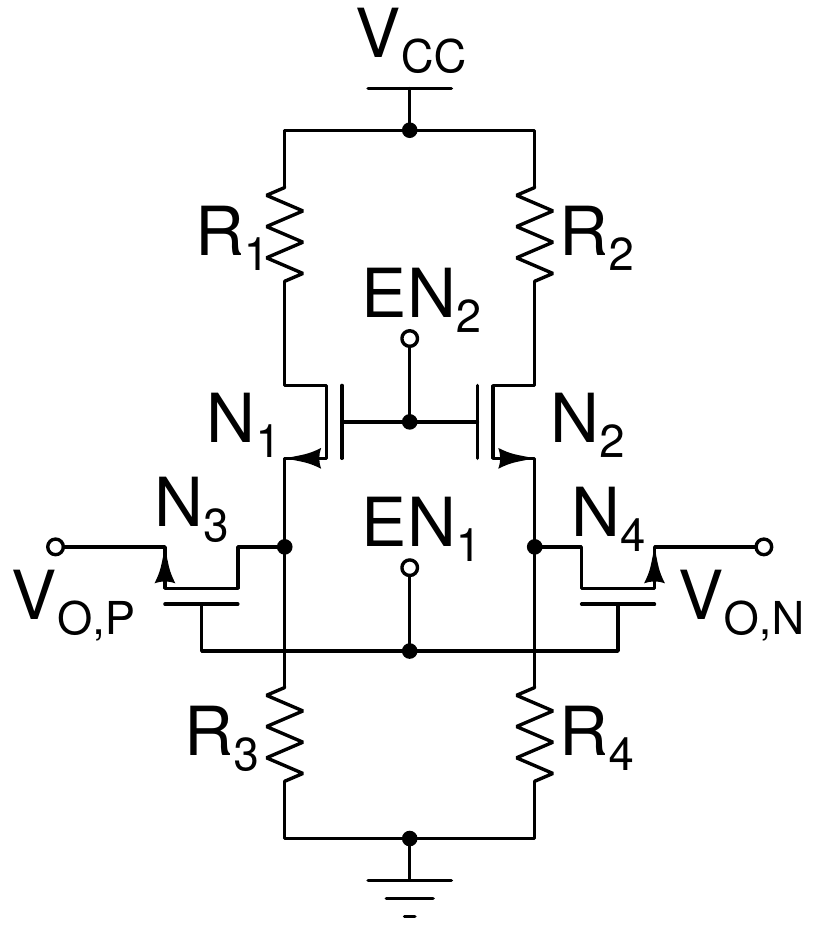} \\[-0.4cm]
      (a) & (b)
    \end{tabular}    
    \caption{Integrator: (a) Schematic. ${E_{L,Q_{P1,2}}}$ = 0.6\,\si{\micro\meter}, ${(W/L)_{N1,2}}$ = 9.52/3, ${(W/L)_{N3,4}}$ = 9.92/3 ${V_{B1}}$ = 0.69\,V ${V_{B2}}$ = 0.7\,V, ${E_{L,Q_{N1,2}}}$ = 0.6\,\si{\micro\meter}, ${(W/L)_{P1,2}}$ = 141.84/3 ${(W/L)_{P3,4}}$ = 72/3, ${V_{B3}}$ = 1.4\,V ${V_{B4}}$ = 1.1\,V, ${(W/L)_{N5,6}}$ = 46/3, ${(W/L)_{N7}}$ = 3.86/0.84, ${(W/L)_{N8,9}}$ = 1.93/0.84, ${V_{REF}}$ = 1.3\,V, ${I_{N1}}$ = 20\,\si{\micro \ampere}, ${(W/L)_{P5}}$ = 145.5/3, ${I_{N6}}$ = 40\,\si{\micro \ampere}, and ${C_L}$ = 7\,pF. (b) Reset circuit connected to the output of the integrator. ${(W/L)_{N1,2}}$ = 20/0.28, ${(W/L)_{N3,4}}$ = 10/0.28, ${R_1+R_3}$ = 1.4\,k$\Omega$, and ${R_2+R_4}$ = 1.4\,k$\Omega$.}   
    \label{fig:integrator_schem}
  \end{figure} 
  
\subsubsection{Integrator}
  
  The settling behavior and steady-state error performance of the equalizer are primarily decided by the DC gain and cut-off frequency of the integrator, which makes it a critical building block. A folded BiCMOS amplifier topology with a large capacitive load is chosen to get a very high DC gain without compromising the pole locations. A circuit schematic of the integrator is shown in Fig. \ref{fig:integrator_schem}(a). The high gain amplifier of the integrator is a modified version of the basic folded cascode amplifier discussed in \cite{R2002} and the G$_{\text m}$ stage of the amplifier is designed with bipolar transistors as suggested in \cite{RK2006}. To obtain a very low-frequency pole without affecting the high gain, the output resistance of the circuit has to be maximized. Hence, the cascode section of the amplifier is biased at a low current. The amplifier gain is given by $|A_V| \approx g_{m,Q_{P1}}\{[(g_{m,P_3}+g_{mb,P_3})r_{o,P_3}(r_{o,Q_{P1}}||r_{o,P_1}) || [(g_{m,N_3}+g_{mb,N_3})r_{o,N_1}r_{o,N_3}]\},$ the output resistance is given by $ R_{O} \approx [(g_{m,P_3}+g_{mb,P_3})r_{o,P_3}(r_{o,Q_{P1}}||r_{o,P_1})] || [(g_{m,N_3}+g_{mb,N_3})r_{o,N_1}r_{o,N_3}],$ and the cut-off frequency is given by $f_{3\,dB} \approx {1}/{2\pi R_OC_L}$ \cite{R2002}. The integrator shown in Fig. \ref{fig:integrator_schem}(a) has a common-mode feedback (CMFB) structure consisting of transistors N$_5$-N$_9$ and the dummy transistors Q$_{\text P2}$ and Q$_{\text N2}$. A detailed design of the CMFB can be found in \cite{R2002}. The integrator has an area of $\sim$120\,\si{\micro\meter}$\times$160\,\si{\micro\meter} and a DC gain of 103.9\,dB with a bandwidth of 56.2\,Hz.
  
  To initialize the weight coefficients, a reset circuit shown in Fig. \ref{fig:integrator_schem}(b) is connected to the output of the integrator. An enable signal, EN$_1$ is applied at the startup of the equalizer operation which turns switches N$_3$ and N$_4$ on. The enable signal EN$_2$, a delayed version of EN$_1$, is used to avoid any unwanted discharge of C$_{\text L}$. Different combinations of R$_1$ to R$_4$ are chosen so as to initialize the weight coefficients at the integrator output nodes. When a global reset signal is asserted the equalizer tap coefficients are reset to the values $\bm h_{xx} = [1+j~~0]^T$, $\bm h_{xy} = [0~~0]^T$, $\bm h_{yx} = [0~~0]^T$, and $\bm h_{yy} = [1+j~~0]^T$.
  
  \begin{figure}[t!]
    \centering
    \begin{tabular}{ll} 
      \includegraphics[scale=0.35]{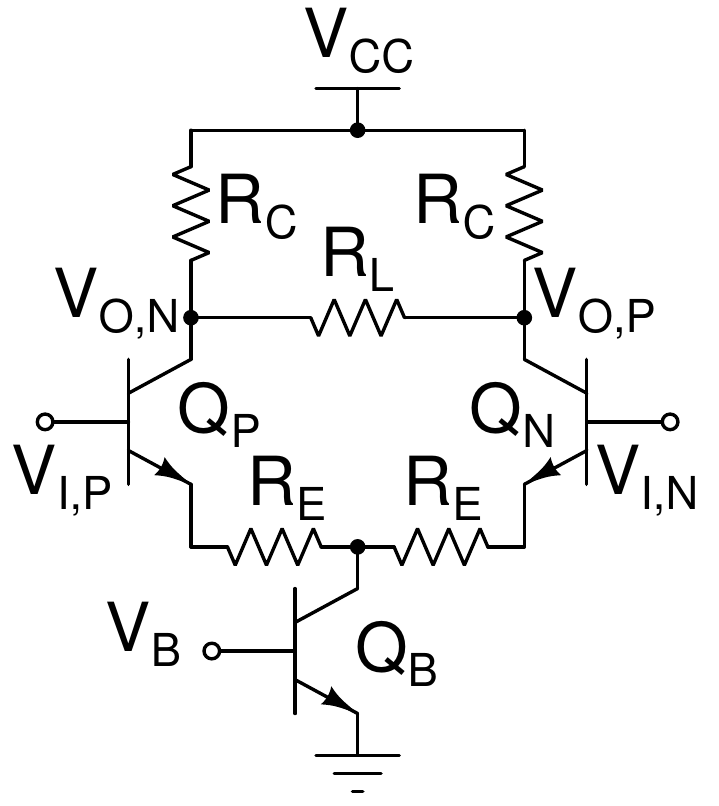} & \includegraphics[scale=0.45]{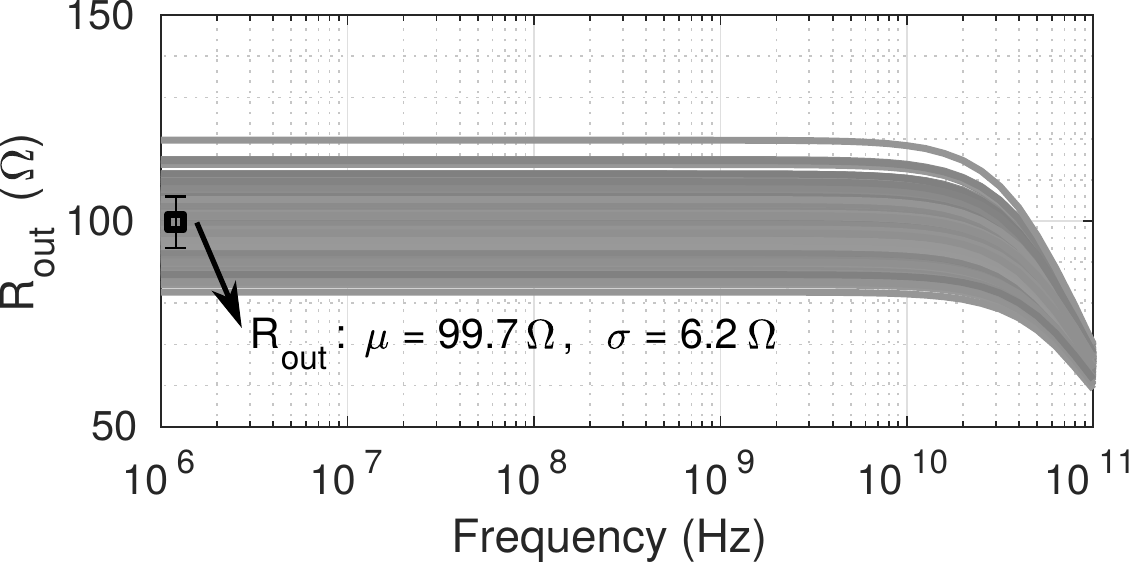} \\[-0.4cm]
      (a) & (b)
    \end{tabular}    
    \caption{Output buffer: (a) Schematic. ${E_{L,Q_{P,N}}}$ = 6\,\si{\micro\meter}, ${E_{L,Q_B}}$ = 6\,\si{\micro\meter}, ${R_C}$ = 320\,$\Omega$, ${R_L}$ = 118\,$\Omega$, ${R_{E}}$ = 12\,$\Omega$, ${V_B}$ = 0.87\,V, and ${I_{N_B}}$ = 6\,mA. (b) Output resistance.}   
    \label{fig:buf_sch}
  \end{figure} 

\subsubsection{Output Buffer}

  Output nodes of the equalizer are designed to drive high-speed transmission lines. To match impedance at the output nodes CE buffers are used, a schematic of which is shown in Fig. \ref{fig:buf_sch}(a). The parallel combination 2R$_{\text C}||$R$_{\text L}$ helps to match the impedance without decreasing the value of R$_{\text C}$. This configuration also helps to achieve different AC and DC output resistances while maintaining the transistors in the active region of operation. The buffer has 4.9\,dB DC gain, 37.8\,GHz bandwidth, and occupies $\sim$44\,\si{\micro\meter}$\times$46\,\si{\micro\meter} area. The output of the buffer has a differential output resistance of 99.7\,$\Omega$ as shown in Fig. \ref{fig:buf_sch}(b).

\subsubsection{Other Building Blocks}

\begin{figure}[t!]
    \centering
    \begin{tabular}{ll}
       \includegraphics[scale=0.35]{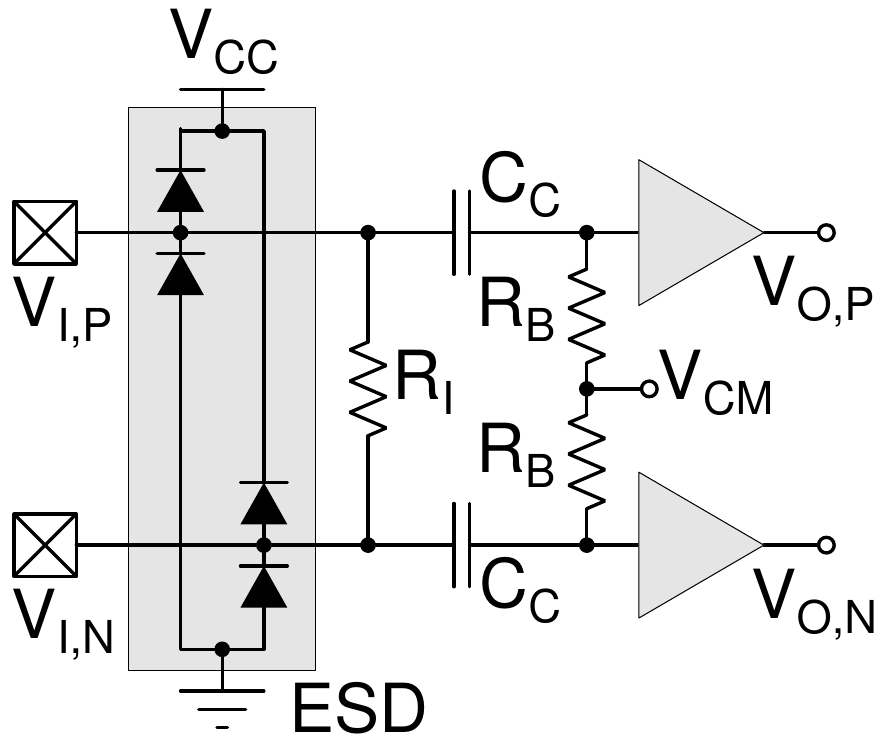} & \includegraphics[scale=0.45]{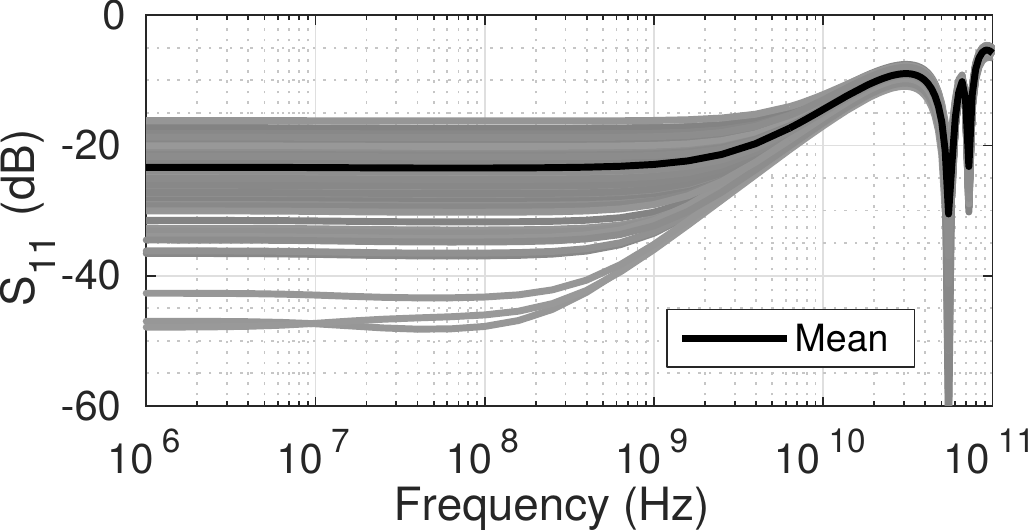} \\[-0.4cm]
      (a) & (b)
    \end{tabular}     
    \caption{Input matching circuit: (a) Schematic. ESD: electro-static discharge protection circuit. ${R_I}$ = 100\,$\Omega$, ${C_C}$ = 500\,fF, ${R_B}$ = 20\,k$\Omega$, and ${V_{CM}}$ = 2.2\,V. (b) Input reflection coefficient of the matching circuit.} 
    \label{fig:in_sch}   
  \end{figure}

  Level shifters are used to shift common-modes up or down at various nodes, and AC coupling is used wherever level shifting is not possible. All bias currents are mirrored from a single source which is supplied from outside. To match the impedance of high-speed inputs, a circuit shown in Fig. \ref{fig:in_sch}(a) is used. This circuit provides 100\,$\Omega$ differential input resistance, and the desired common mode to the input signals. There is also an electro-static discharge (ESD) protection circuit made up of ESD diodes. Simulations show an $S_{11}$ parameter, which is better than \textminus10\,dB in the frequency band of concern, as shown in Fig. \ref{fig:in_sch}(b).
  
\subsection{The Equalizer IC}  

  A micrograph of the prototype equalizer is shown in Fig. \ref{fig:cma_eq}(c). The IC has 50 pads of which 28 are meant for high-speed differential signals which are arranged in a ground-signal-signal-ground-signal-signal-ground pattern. Rest of the pads are used for reset signal, bias current, amplitude control signals, and supply voltage and ground. The equalizer occupies $\sim$1.4\,mm$\times$1.35\,mm chip area and consumes 2.5\,W. The power consumption remains the same for data rates of up to 100\,Gb/s since the chip uses a constant current biasing.

\section{Results and Discussion}
\label{sec:results}
  
  The CMA equalizer is designed to operate at 100\,Gb/s. However, high-speed assembly/packaging constraints limited testability of the chip to 40\,Gb/s data rate.

\subsection{Measurement Results at 40\,Gb/s Data Rate}
\label{subsec:exp}
  
  \begin{figure*}[t!]
    \centering
    \includegraphics[scale=0.82]{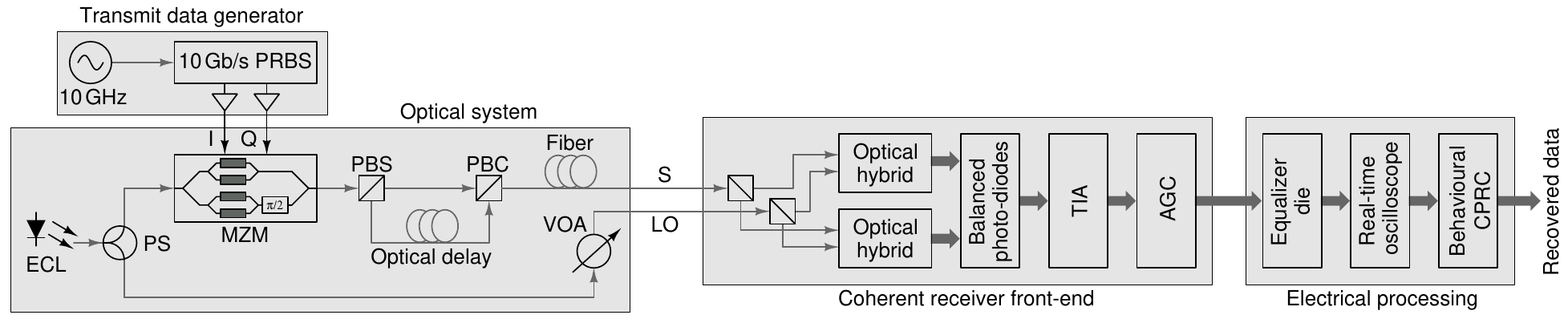}
    \caption{Experimental setup with a 40\,Gb/s DP-QPSK optical transmission system. PRBS: pseudo random binary sequence, I: in-phase data, Q: quadrature-phase data, ECL: external cavity laser, PS: power splitter, MZM: Mach-Zehnder modulator, PBS: polarization beam splitter, PBC: polarization beam combiner, VOA: variable optical attenuator, S: optical signal from the fiber channel, LO: local oscillator, TIA: trans-impedance amplifier, AGC: automatic gain control, and CPRC: carrier phase recovery and compensation.} 
    \label{fig:setup}
  \end{figure*} 

  Fig. \ref{fig:setup} shows a block diagram of the experimental setup with a 40\,Gb/s DP-QPSK system. In this setup, an external cavity laser of 1550\,nm wavelength is used as the carrier source at the transmitter. A 50:50 power splitter (PS) divides the laser output into two parts--one of which gets modulated at the transmitter and the other is used as a local oscillator (LO) at the receiver. Two independent 10 Gb/s data streams are generated using a 10 Gb/s pseudo-random binary sequence generator which is clocked by a 10\,GHz source. The two data streams are amplified to drive the nested Mach-Zehnder modulator (MZM). The MZM gives out a 20 Gb/s QPSK modulated carrier which is split into X and Y polarizations using a polarization beam splitter (PBS). The X output of the PBS is directly connected to the X input of a polarization beam combiner (PBC), whereas the Y output is delayed using a 2\,m-long polarization maintaining fiber before connecting to the Y input of the PBC. This optical delay is used to de-correlate the X and Y QPSK signals. The MZM-PBS-optical delay-PBC combination emulates a DP-QPSK modulator which gives out a 40\,Gb/s DP-QPSK modulated carrier to the channel.
   
  \begin{figure}[t!]
    \centering 
    \includegraphics[scale=0.07]{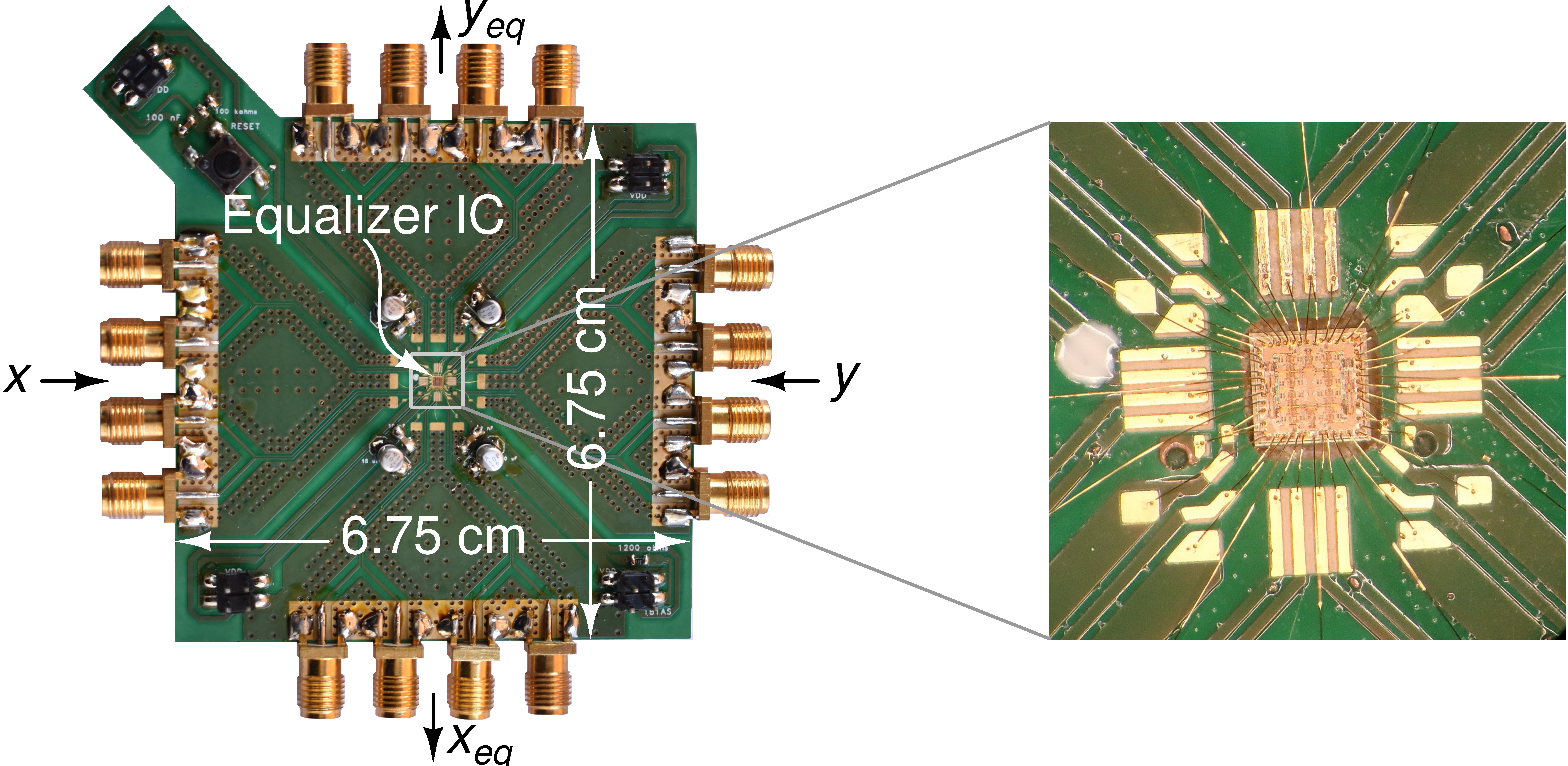}
    \caption{PCB used to validate the CMA Equalizer. The equalizer IC is wire-bonded to the center of the PCB as shown in the inset.} 
    \label{fig:board}
  \end{figure} 
  
  \begin{figure}[t!]
    \centering
    \includegraphics[scale=0.75]{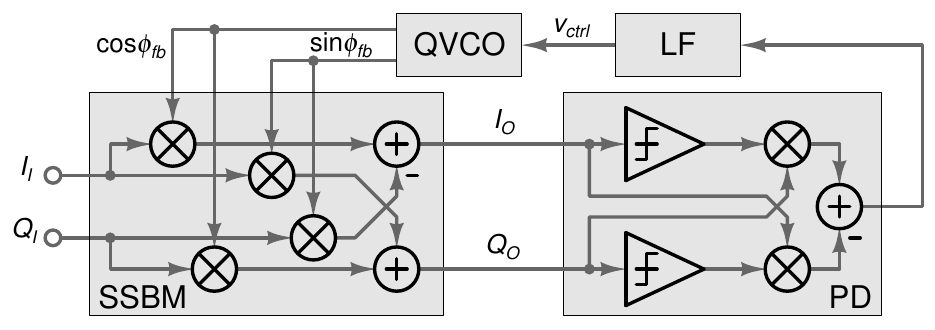} 
    \caption{Architecture of the decision-assisted Costas loop CPRC \cite{NN2015}. SSBM: single sideband mixer, PD: phase detector, LF: loop filter, and QVCO: quadrature-phase voltage controlled oscillator.} 
    \label{fig:cprc}   
  \end{figure}
    
  At the receiver side, the signal from the channel, S and the LO are fed to the inputs of an integrated coherent optical receiver front-end. The required power level of the LO is maintained using a variable optical attenuator (VOA), which is connected just before the receiver front-end. The receiver front-end consists of PBSs, 90\textdegree optical hybrids, balanced photo-diodes, trans-impedance amplifiers, and automatic gain control in the mentioned order. This module maintains a differential signal level of 400\,mVpp on all four inputs of the equalizer IC which is wire-bonded to a printed circuit board (PCB). The PCB shown in Fig. \ref{fig:board} is fabricated using RT duriod/6010LM laminate, more details of which can be found in \cite{TN2019}. The PCB has four pairs of input transmission lines as well as output transmission lines which end on SMA connectors mounted along the periphery of the PCB. The outputs of the equalizer are stored using a 21\,GHz real-time oscilloscope for further processing with a behavioral model of a CPRC. Fig. \ref{fig:cprc} shows the block diagram of the behavioral CPRC that consists of a single sideband mixer, phase detector, loop filter, and quadrature-phase voltage controlled oscillator. Further details of the CPRC can be found in \cite{NN2015}. 
  
  The error vector magnitude (EVM) of the output signal is calculated by sampling the CPRC outputs at the maximum eye-opening point. The EVM is calculated as $$EVM (\%) = \frac{\sqrt{\frac{1}{N}\sum\limits_{k=1}^N\left\{(I_k-\widehat{I_k})^2+(Q_k-\widehat{Q_k})^2\right\}}}{A}\times 100\%$$ where $I_k$ and $Q_k$, respectively are the I and Q components of the $k^{\text{th}}$ received symbol, $\widehat{I_k}$ and $\widehat{Q_k}$, respectively are the expected I and Q components of the $k^{\text{th}}$ received symbol, $A$ is the expected radius of the constellation points, and $N$ is the total number of received symbols.
  
\begin{figure}[t!]
    \centering
    \begin{tabular}{ccc}
      \includegraphics[scale=0.155]{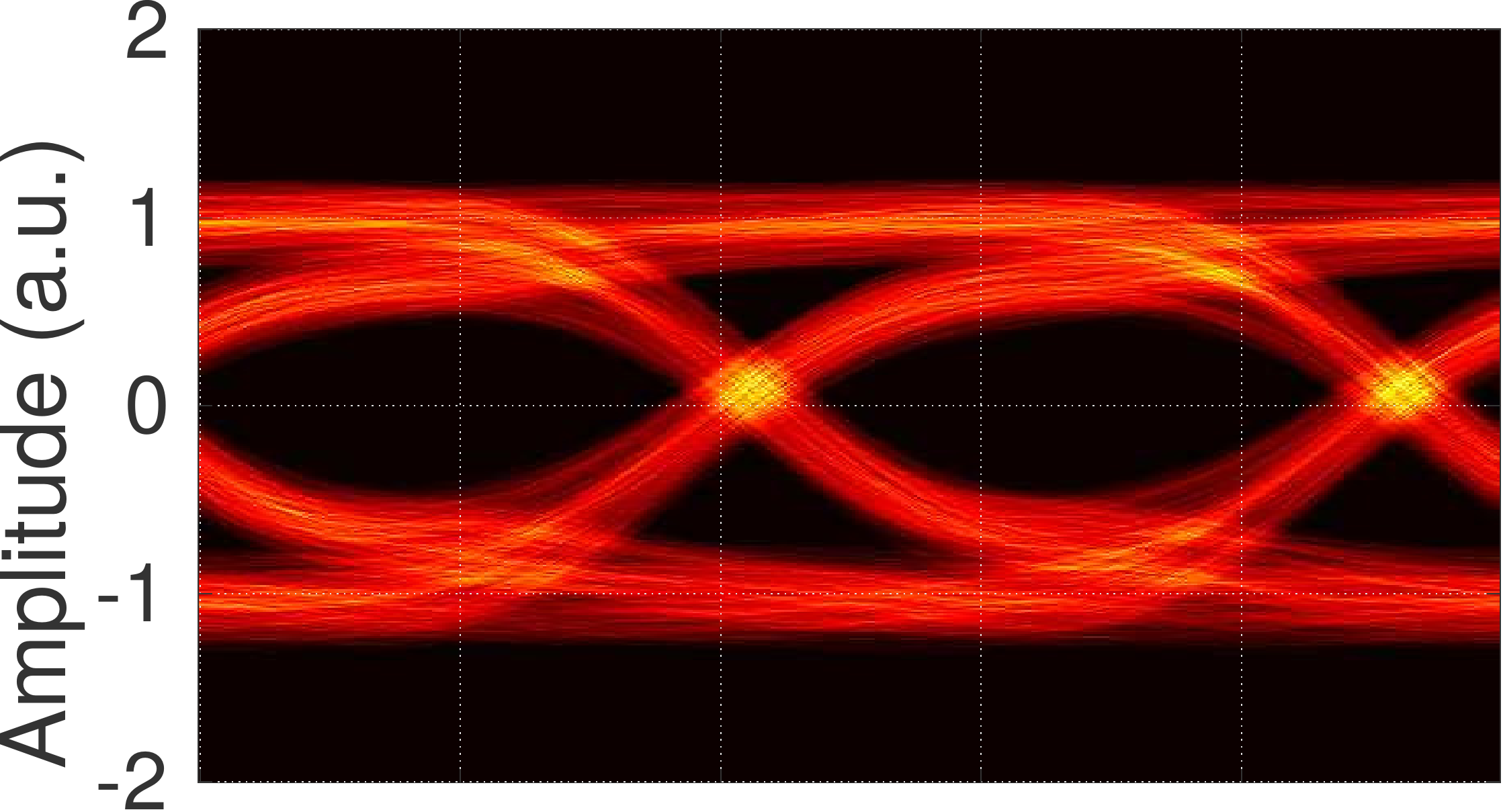} & \hspace{-0.25cm} &
      \raisebox{0.04\height}{\hspace{0.1cm}\includegraphics[scale=0.155]{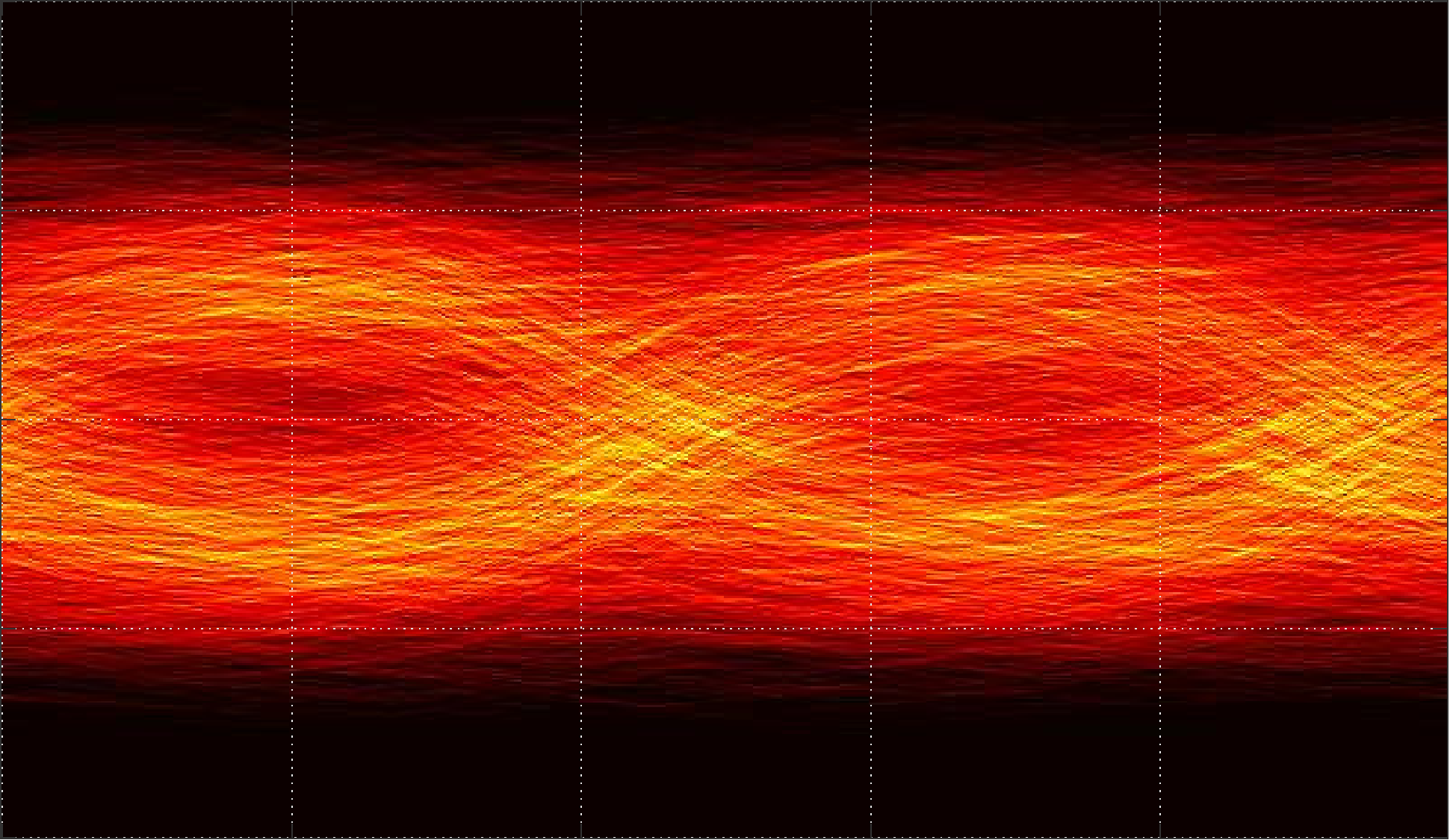}} \\[-17pt] 
      \hspace{-2cm}\color{white}(a) & \hspace{-0.25cm} & \hspace{-2.4cm}\color{white}(b) \\ [6pt]
      \includegraphics[scale=0.155]{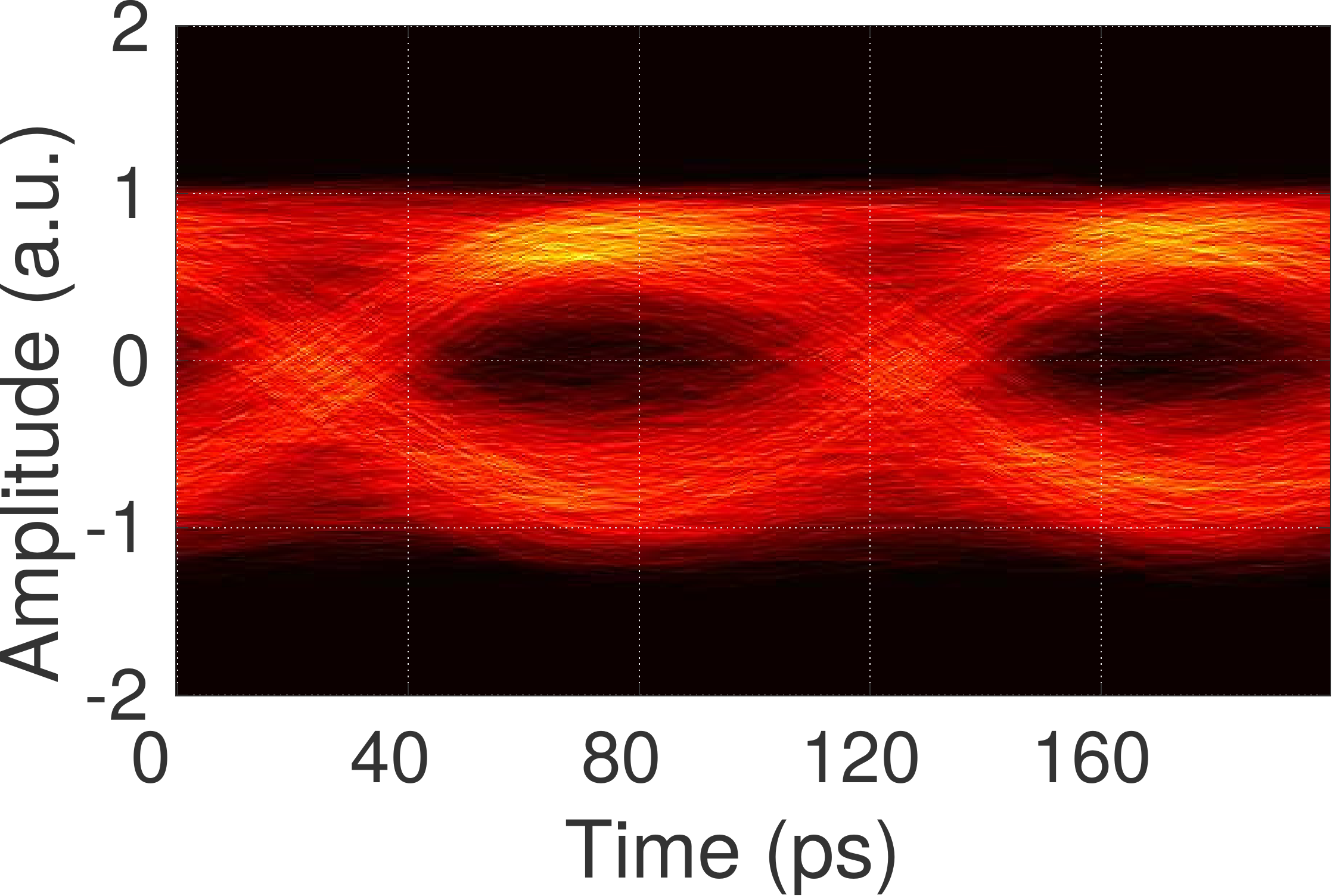} & \hspace{-0.25cm} &
      \includegraphics[scale=0.155]{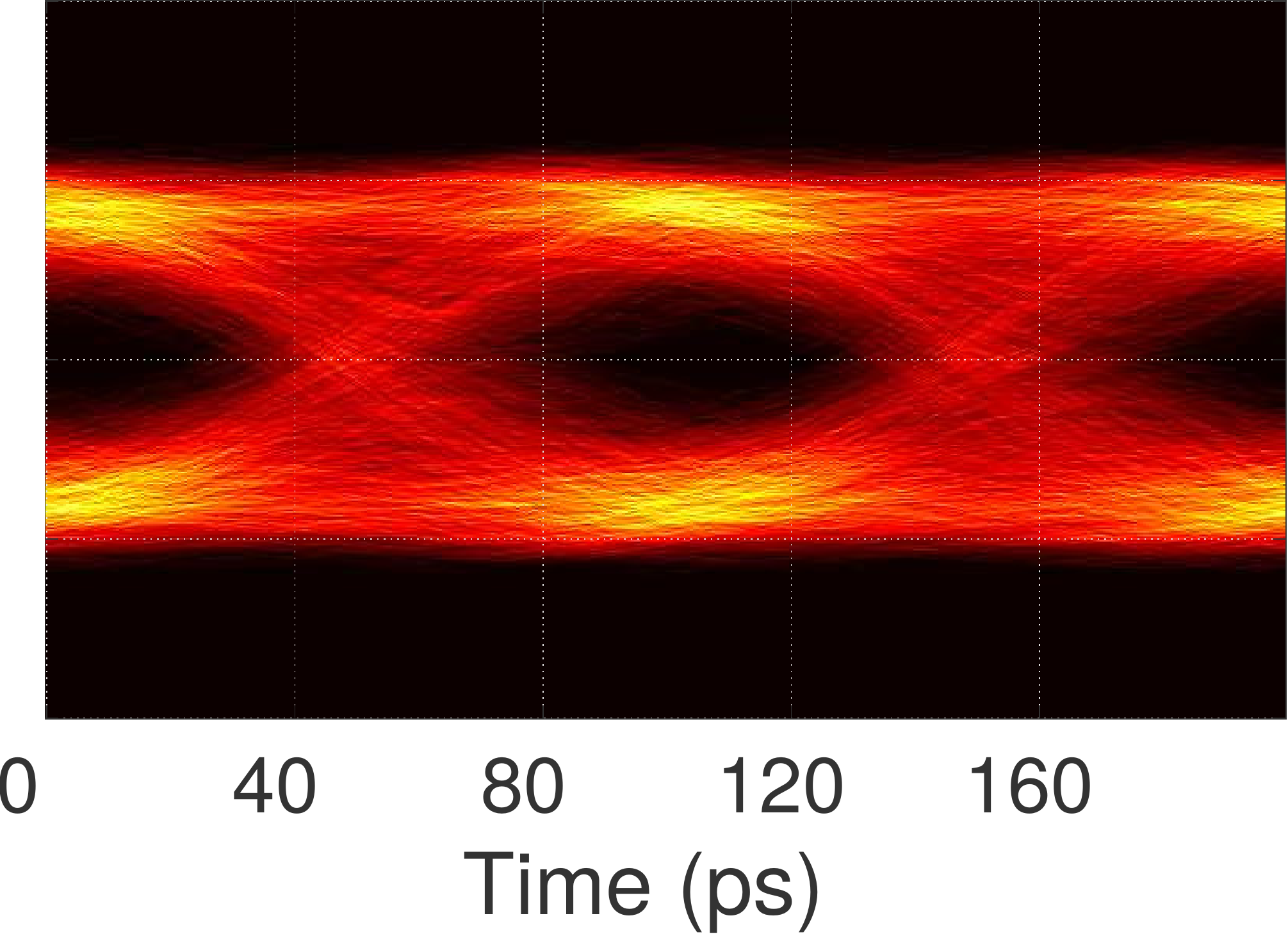} \\[-31pt] 
      \hspace{-2cm}\color{white}(c) & \hspace{-0.25cm} & \hspace{-2.4cm}\color{white}(d)\\ [12pt]
    \end{tabular}
    \caption{Measurement results: eye-diagrams of the in-phase component of X polarization signal at the (a)  transmitter, (b) equalizer chip's input, (c) equalizer chip's output, and (d) CPRC output in the experimental setup with a 40\,Gb/s back-to-back link.}
    \label{fig:expeyes}   
  \end{figure}

  Performance of the equalizer is characterized using the 40\,Gb/s system for different link lengths. Fig. \ref{fig:expeyes} shows the eye-diagrams obtained at various stages of the experimental setup with a back-to-back optical link. The eye-diagram shown in Fig. \ref{fig:expeyes}(a) is that of the transmitted signal with 24\% EVM. No pre-equalization or pre-compensation are performed to nullify the effects of components non-idealities, which resulted in a poor transmitter EVM. Fig. \ref{fig:expeyes}(b) shows the received signal eye-diagram, which is distorted due to the system non-idealities. Fig. \ref{fig:expeyes}(c) shows the equalizer chip's output eye-diagram and Fig. \ref{fig:expeyes}(d) shows the behavioral CPRC's output eye-diagram with 28\% EVM which corresponds to an estimated bit error rate (BER) of $1.8\times10^{-4}$. The BER is estimated from the EVM as $$ BER \approx \frac{2(1-1/P)}{\log_2P}Q\left\{\sqrt{\bigg(\frac{3\log_2P}{P^2-1}\bigg)\bigg(\frac{2}{EVM^2\log_2M}\bigg)}\right\}$$ where $P$ is number of levels per dimension and $M$ is the number of symbols on the I-Q plane \cite{SR2006}.
    
  \begin{figure}[t!]
    \centering
    \begin{tabular}{ccc}
      \includegraphics[width=2.4cm]{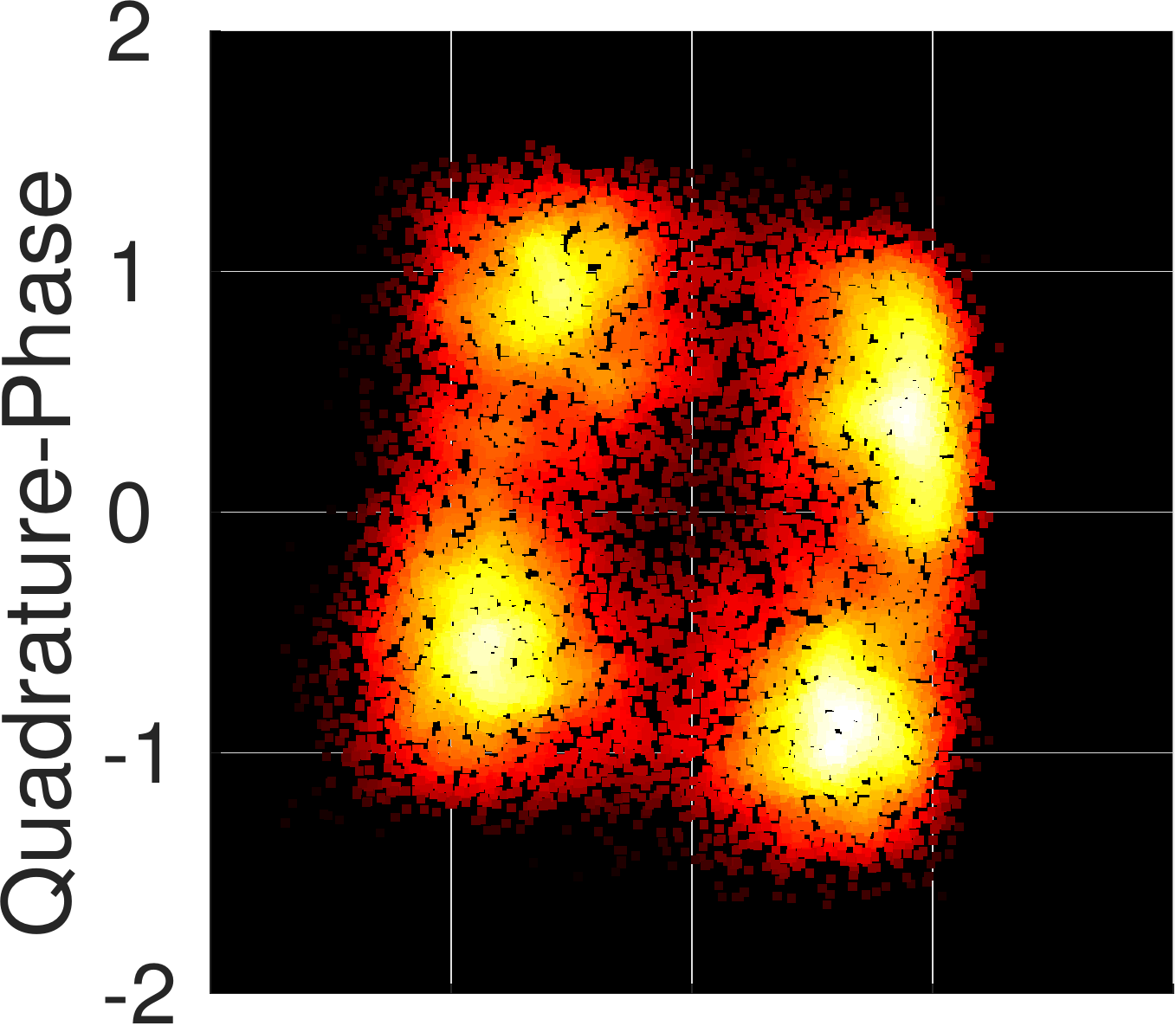} &
      \raisebox{0.02\height}{\hspace{0.1cm}\includegraphics[width=2.0cm]{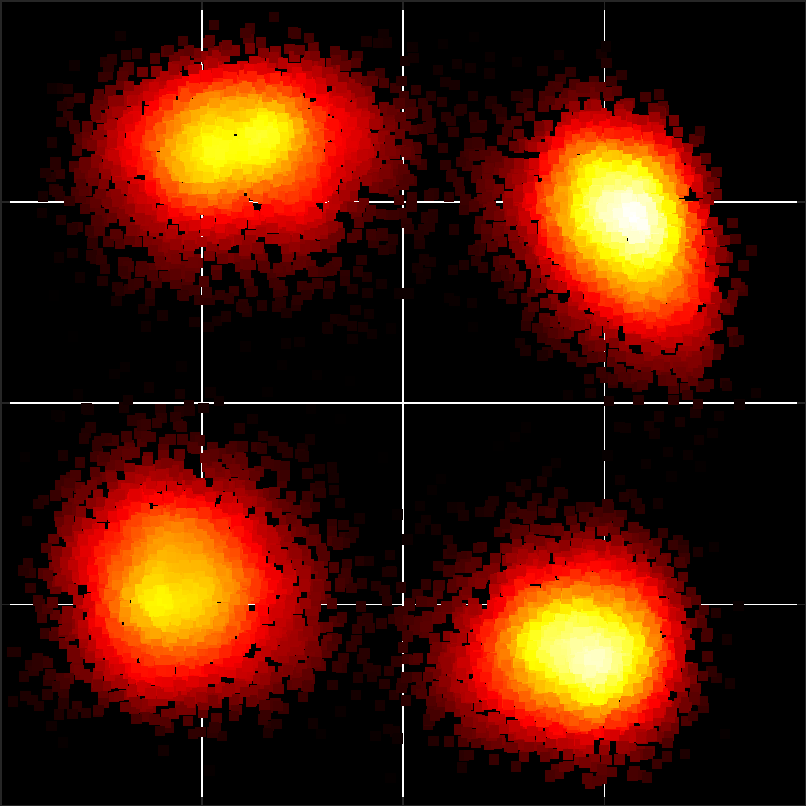}} &
      \raisebox{0.02\height}{\hspace{0.1cm}\includegraphics[width=2.0cm]{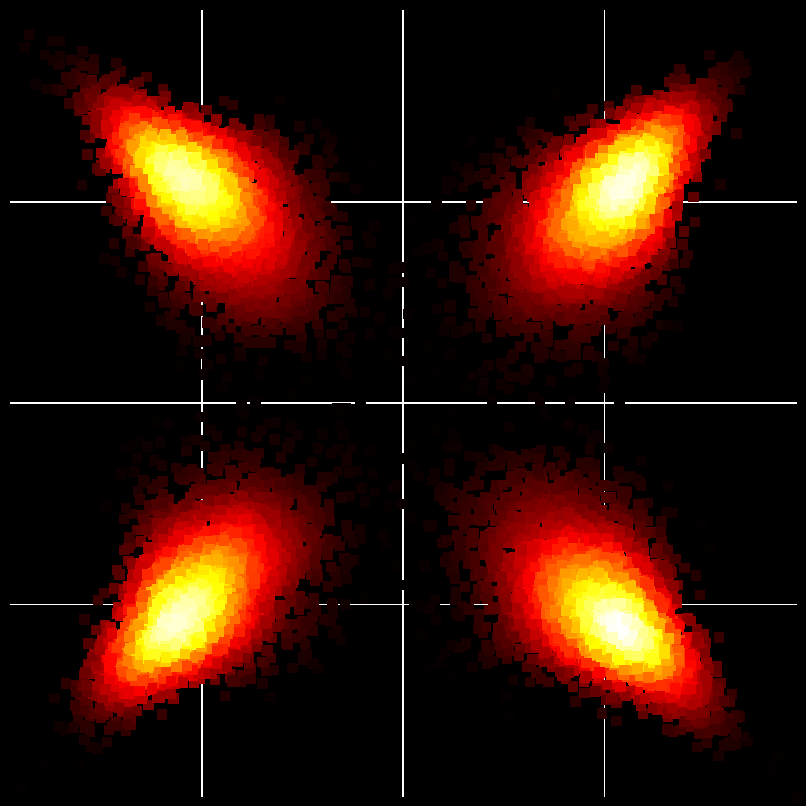}} \\[4pt]
      \includegraphics[width=2.4cm]{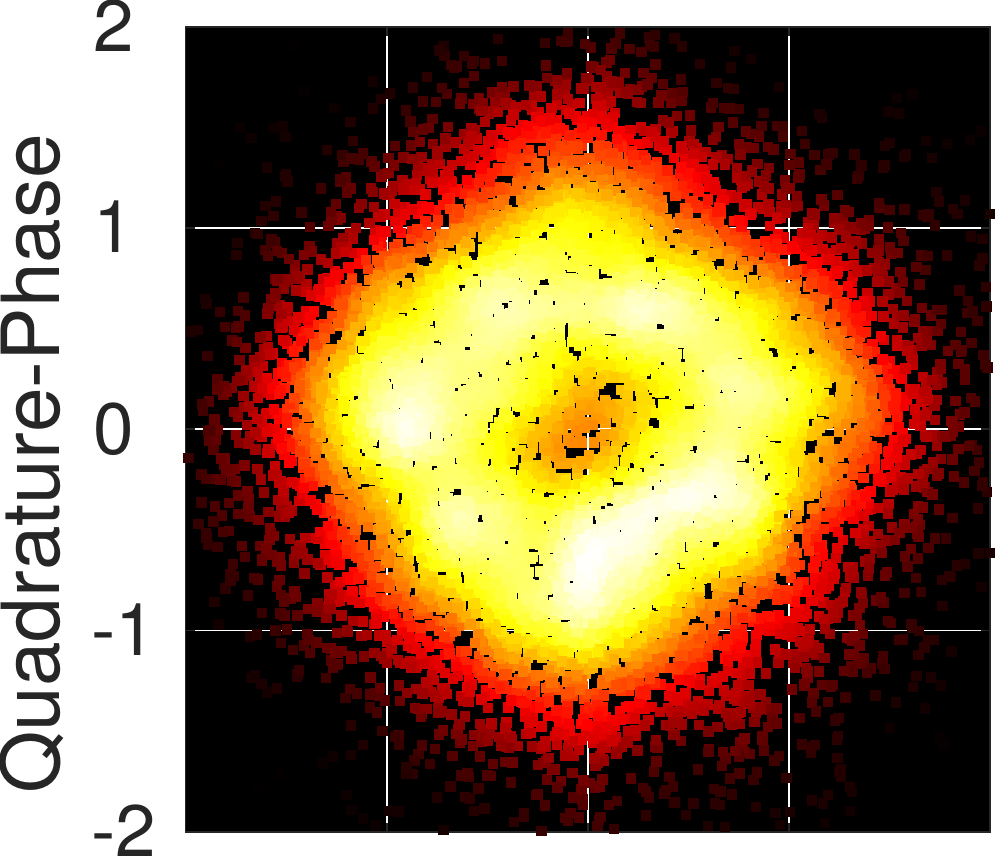} &
      \raisebox{0.02\height}{\hspace{0.1cm}\includegraphics[width=2.0cm]{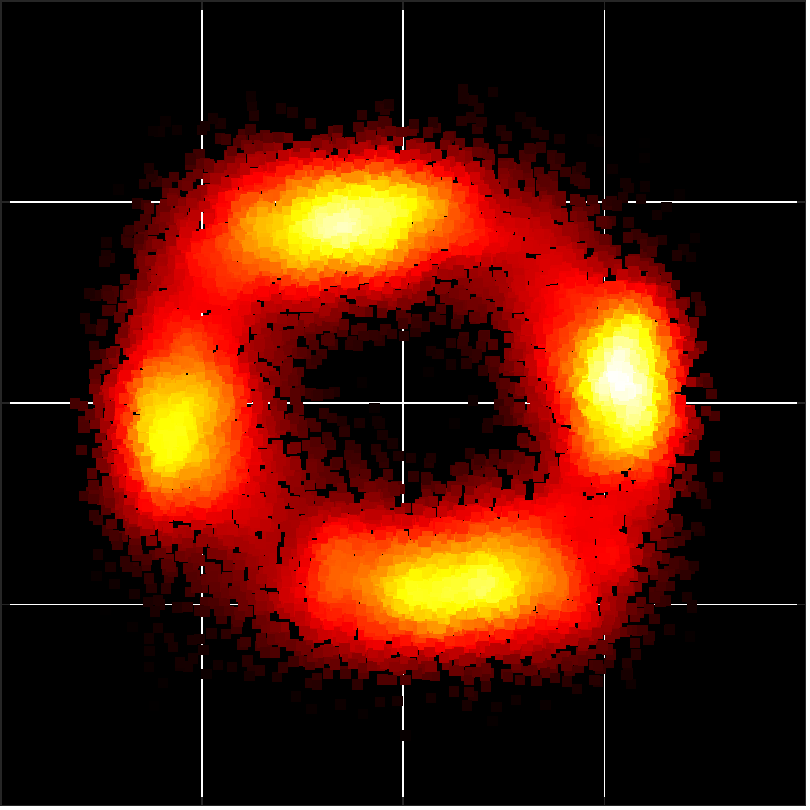}} &
      \raisebox{0.02\height}{\hspace{0.1cm}\includegraphics[width=2.0cm]{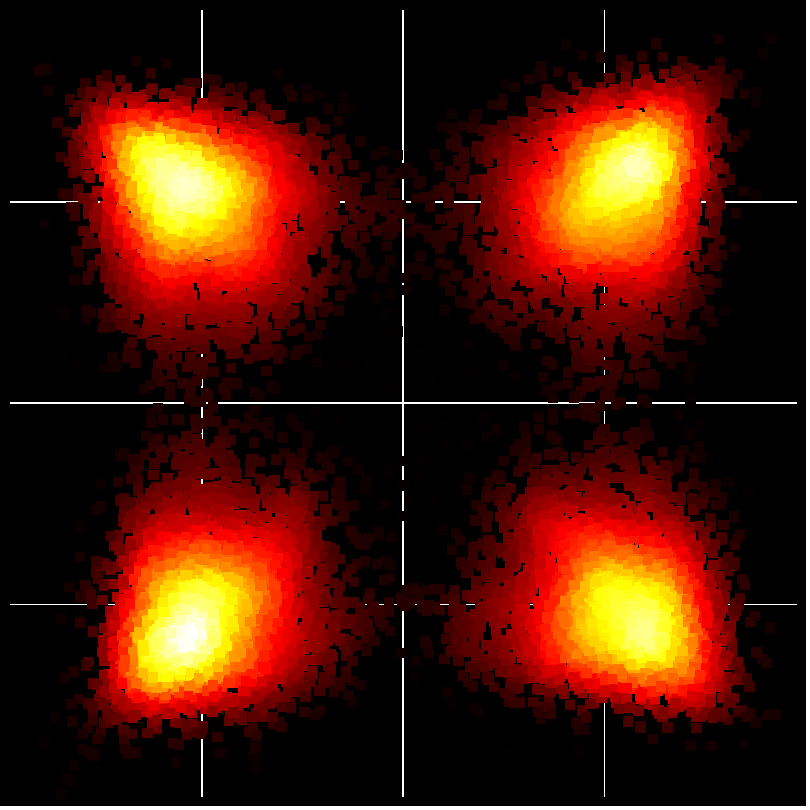}} \\[4pt]
      \includegraphics[width=2.4cm]{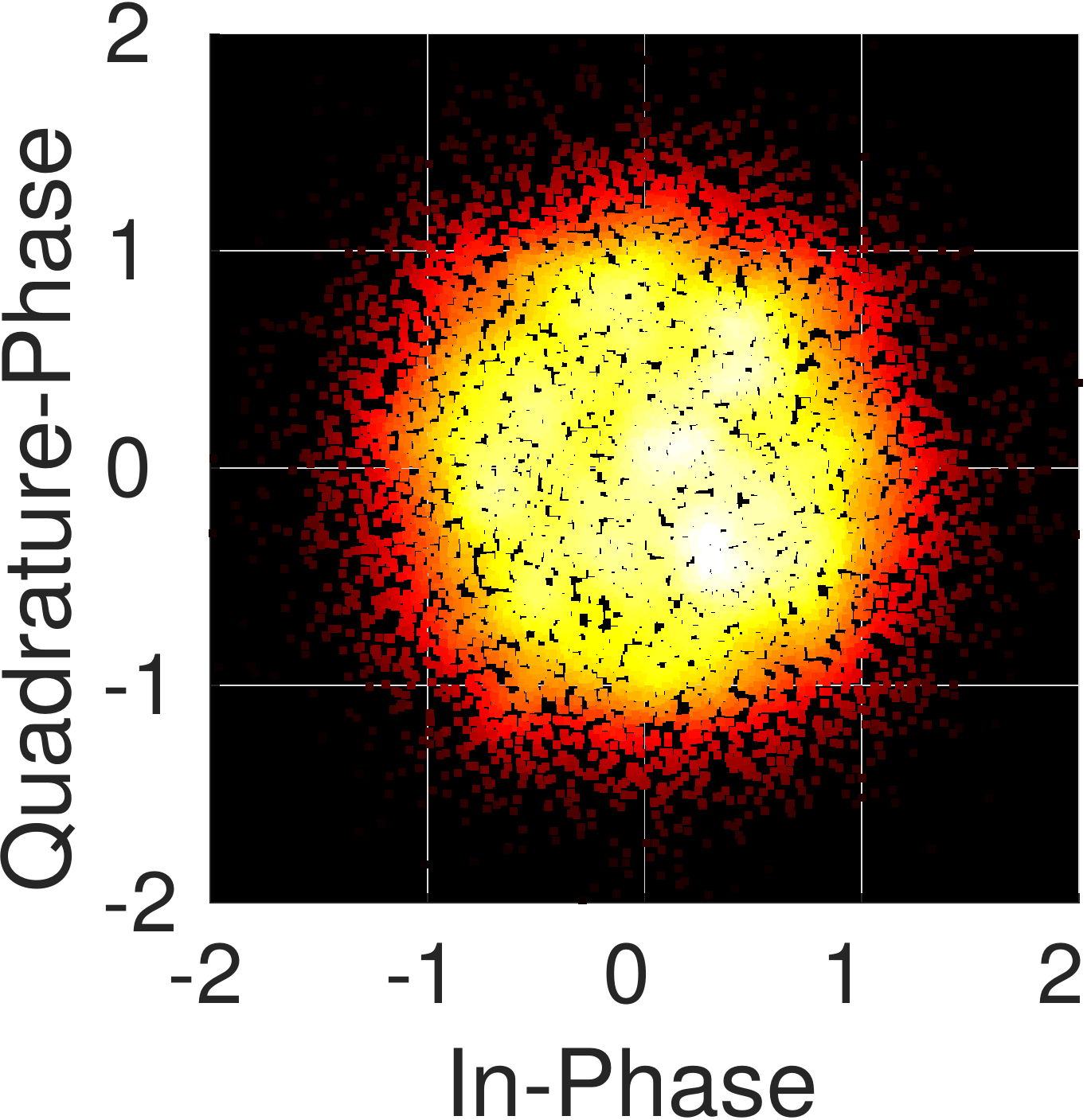} &
      \includegraphics[width=2.02cm]{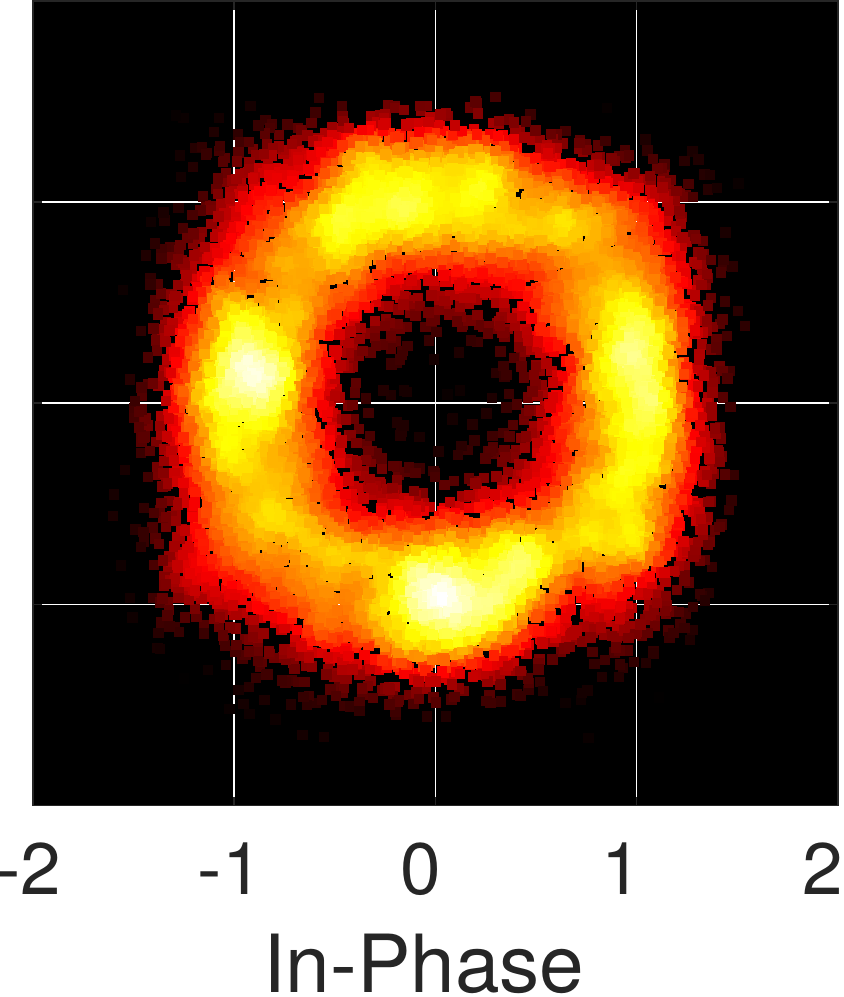} &
      \includegraphics[width=2.02cm]{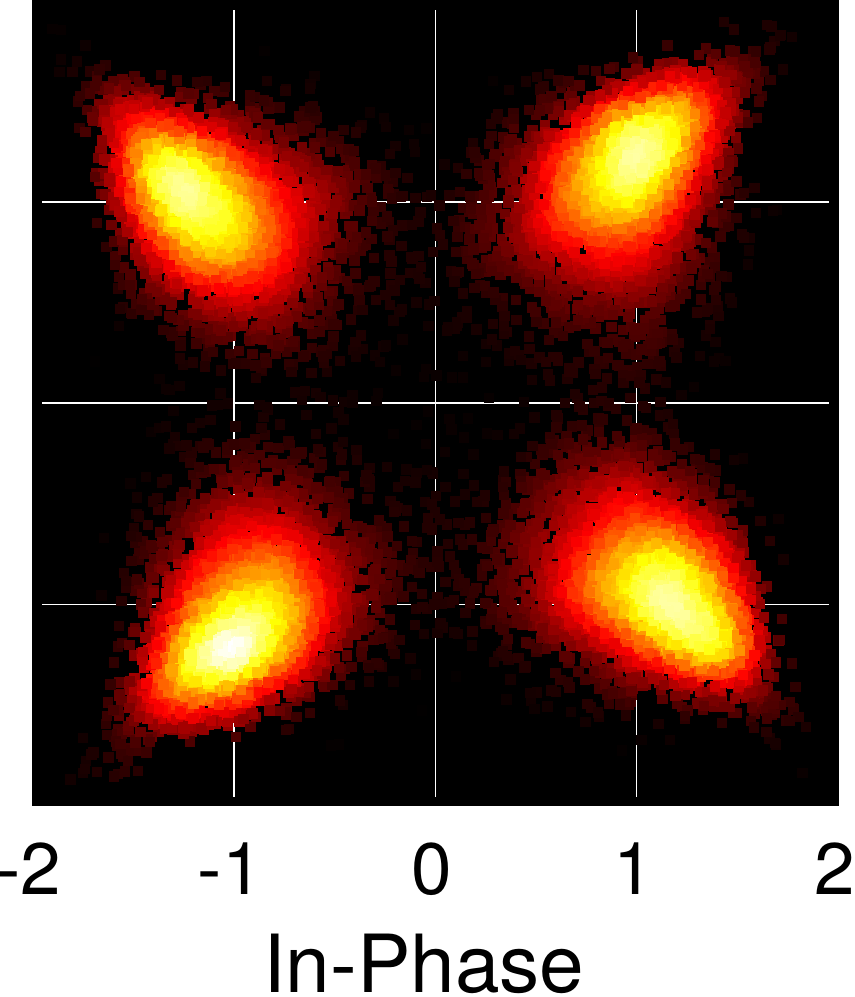}\\
      (a) & (b) & (c) 
    \end{tabular}
    \caption{Measurement results: X polarization constellations of the (a) received signal, (b) equalizer chip's output signal, and (c) recovered signal in a 40\,Gb/s back-to-back (top row), 5\,km (middle row), and 10\,km (bottom row) links.} 
    \label{fig:expconst}   
  \end{figure}
  
  X polarization constellations at various stages of the system are shown in Fig. \ref{fig:expconst} with results of a back-to-back link in the top row, a 5\,km link in the middle row, and a 10\,km link in the bottom row. The received optical powers are \textminus7.2\,dBm, \textminus10.4\,dBm, and \textminus11.6\,dBm, respectively and the minimum requirement of the coherent receiver front-end is \textminus20\,dBm. Since the same laser is used at the transmitter and receiver, there is only a minimal phase offset between the received signal and LO at the receiver in the back-to-back link. Under this condition, the equalizer recovers four constellation points, but with a slight rotation due to the phase insensitivity of the CMA algorithm. However, in 5\,km and 10\,km links, a large optical path difference between the signal and LO results in a frequency offset which is evident from the equalized constellation diagrams, which appear as rings on the I-Q plane. Post-processing carried out with the behavioral CPRC results in the constellations shown in Fig. \ref{fig:expconst}(c) with an EVM of 28\% in the back-to-back, 32\% in the 5\,km, and 33\% in the 10\,km links. The corresponding estimated pre-forward error correction (pre-FEC) BERs are $1.8\times10^{-4}$, $8.9\times10^{-4}$, and $1.2\times10^{-3}$, respectively, which are well below the hard-decision FEC limit. The intensity noise in the output constellation can be attributed to the frequency dependent S-parameters of the transmission lines of the test PCB \cite{TN2019}.
  
\subsection{Post Layout Simulation Results at 100\,Gb/s Data Rate}
\label{subsec:sims}

\begin{figure}[t!]
    \centering
    \begin{tabular}{ccc}
      \includegraphics[scale=0.325]{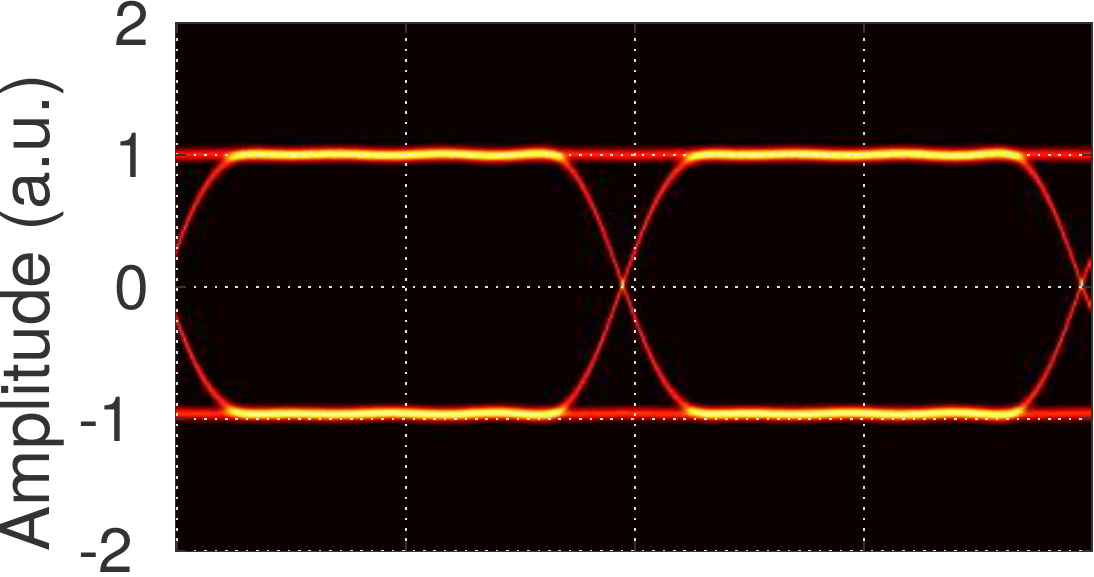} & \hspace{-0.25cm} &
      \includegraphics[scale=0.325]{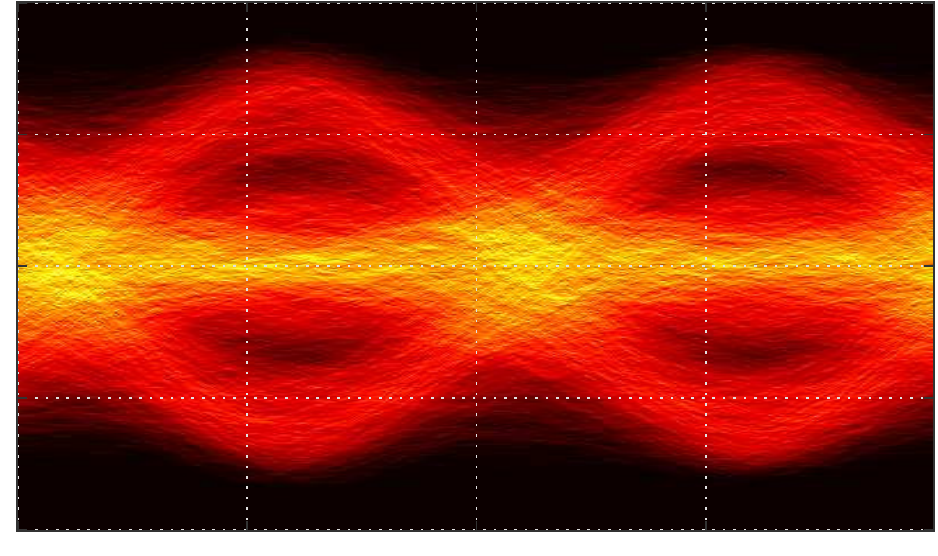} \\[-17pt] 
      \hspace{-1.9cm}\color{white}(a) & \hspace{-0.25cm} & \hspace{-2.4cm}\color{white}(b) \\ [6pt]
      \includegraphics[scale=0.325]{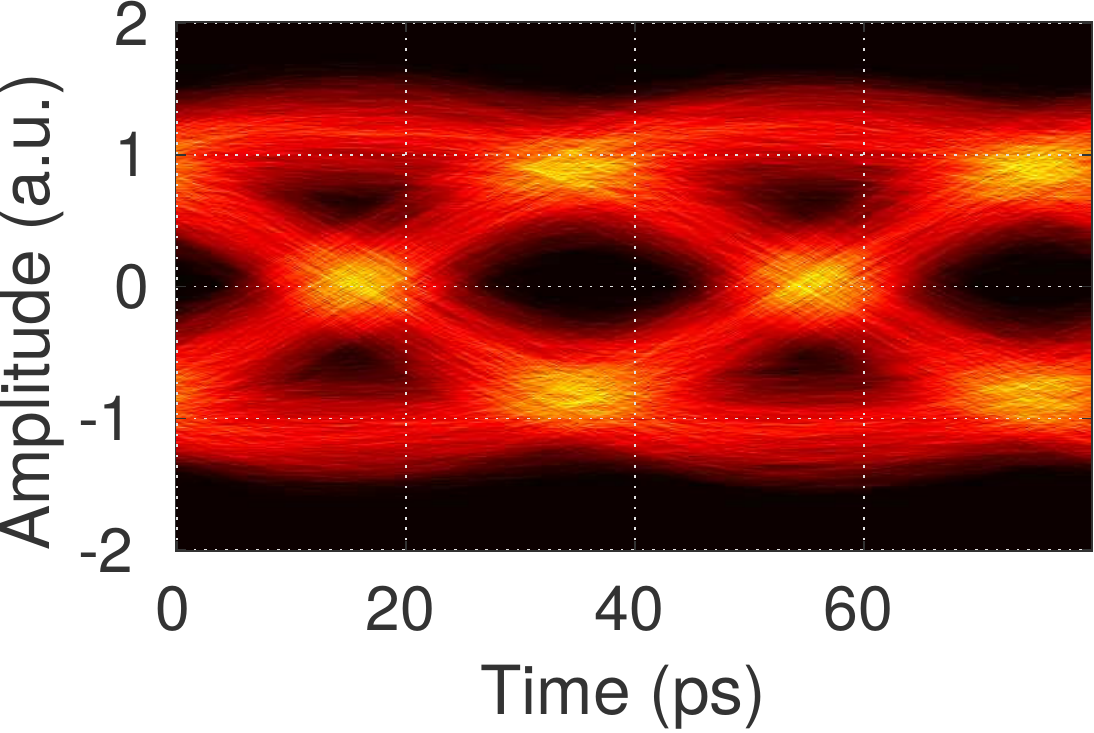} & \hspace{-0.25cm} &
      \includegraphics[scale=0.325]{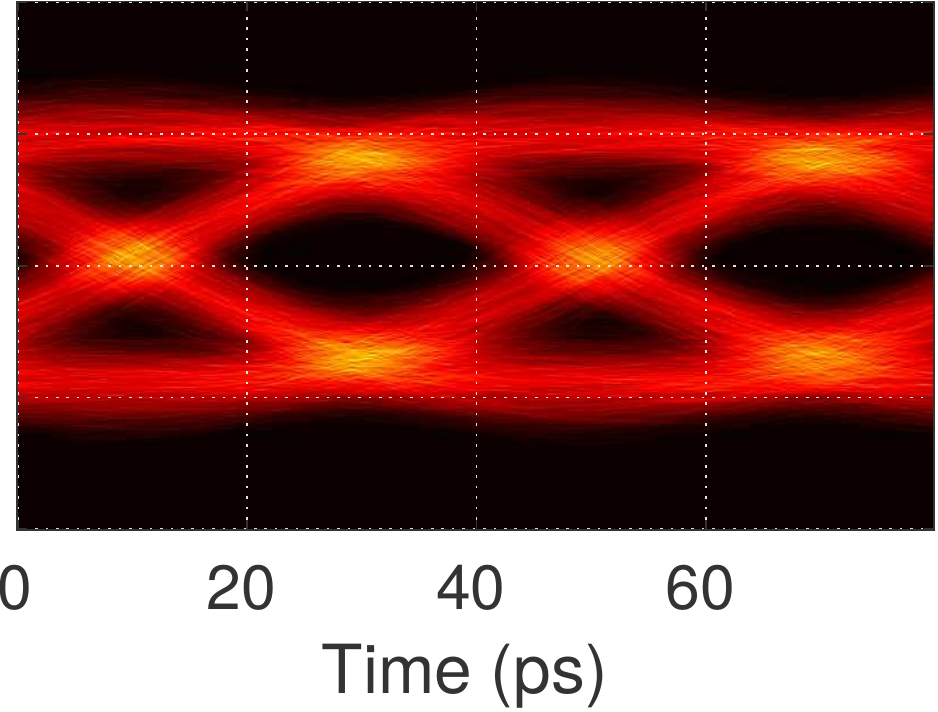}\\[-33pt] 
      \hspace{-1.9cm}\color{white}(c) & \hspace{-0.25cm} & \hspace{-2.4cm}\color{white}(d) \\ [12pt]
    \end{tabular}
    \caption{Post-layout simulation results with 100\,Gb/s signals from a simulation model of 5\,km optical link. Eye-diagrams of the in-phase component of X polarization signals at the (a) transmitter, (b) equalizer input, (c) equalizer output, and (d) CPRC output.} 
    \label{fig:simseyes}   
  \end{figure}

  The small pad-pitch and the large pin count of the chip limited the board assembly and packaging options, hence the experimental validation of the equalizer at the designed data rate is not performed. Also, the small pad pitch of the IC limited the possibilities of an optimal direct die attach and a very well matched on-PCB transmission lines. Apart from these, issues owing to the bandwidth limitation due to bond-wire inductance, transmission lines, and SMA connectors, functionality of the equalizer at 100\,Gb/s is verified through post-layout simulations. A 100\,Gb/s transmission system is modeled in VPItransmissionMaker from which the received signals after the O/E conversion are exported for post-layout circuit simulation. Fig. \ref{fig:simseyes} shows the results of a post-layout simulation carried out in typical-typical corner with the data from a simulation model of an optical link with a 5\,km SMF channel. Fig. \ref{fig:simseyes}(a) shows the eye-diagram of the transmitted data with an EVM of 1\%, which becomes distorted when it reaches the receiver side, as shown in Fig. \ref{fig:simseyes}(b). Fig. \ref{fig:simseyes}(c) is the eye-diagram of the equalizer output and Fig. \ref{fig:simseyes}(d) is the output of a behavioral CPRC with an EVM of 27.8\%.
  
\subsection{Challenges and Future Directions}
  
  Due to packaging limitations, the measured performance of the equalizer IC is sub-optimal and the IC could not be tested beyond 40\,Gb/s data rate. A flip-chip-on-board assembly would have resulted in a better performance from the IC, but such a setup was not feasible as the pad-pitch is very small. A fine-tuned optical system employing pulse shaping at the transmitter side to help with effective bandwidth utilization and a PCB with well-matched transmission lines are also expected to improve the experimental setup. The major limiting factor of all-analog processing would be the bandwidth reduction due to the cascading effect. This would restrict the maximum amount of dispersion to be corrected since a larger amount of dispersion requires a larger number of cascaded taps in the equalizer. However, usage of on-chip transmission lines and a combination of active and passive delay lines would help to increase the number of taps. Also, optimizing the delay cell can increase the number of cascaded stages in the delay line \cite{SG2015, MV2020}. It is seen that the delay cells contribute significantly to the overall power consumption of the IC. However, by optimizing the delay cell in the same node a significant overall power reduction is also reported \cite{AS2016}. 
  
  Usage of a lower technology node will help to reduce the power consumption, which may be inferred from \cite{LF2017, DK2018, LK2018, XD2017}. Implementation using FinFET technology is also expected to reduce power consumption and improve performance because of a lower supply voltage and a higher transit frequency ($f_T$). For example, an ADC, which is an analog circuit, in 14\,nm FinFET technology \cite{LK2018} shows a 30X improvement in energy efficiency (mW/bit/GHz) over an ADC in 130\,nm BiCMOS \cite{XD2017}. If implemented using FinFET Technologies, we can expect the equalizer's energy efficiency (mW/Gb/s) also to improve by a similar factor.
  
  The equalizer presented in this paper is meant for DP-QPSK format. By using the analog blocks designed for the equalizer it is possible to implement other adaptive algorithms too. For example, a 16QAM equalizer that uses a radius directed  equalizer algorithm is presented in \cite{YP2016}. Other critical blocks at the receiver side are also being investigated successfully for 16QAM. For instance, an analog domain Costas loop-based CPRC IC for QPSK and 16QAM formats is demonstrated quite recently in \cite{RA_OFC2019, RA2020}. These advances give a clear indication that the ASP technique is scalable and is a good solution for the power consumption-size-cost problem of the DCI transceivers. A comprehensive comparative study on DSP and ASP techniques is reported in \cite{RK2020}. It shows an energy consumption of 65.53\,mW/Gb/s by an ASP based coherent receiver while that of 210.46\,mW/Gb/s by a DSP based coherent receiver at 200\,Gb/s operation emphasizing a $\sim$3X reduction in energy consumption per bit using ASP.

\section{Conclusion}
\label{sec:conclusion}

  The detailed architecture and demonstration of the first all-analog equalizer for DP-QPSK links are presented in this paper. A two-tap proof-of-concept prototype of the equalizer, that uses the CMA algorithm to update weight coefficients, is implemented in a 130\,nm BiCMOS technology. Measurement results of the equalizer IC obtained with a 40\,Gb/s, back-to-back link show an EVM that is very close to the transmitter EVM, which clearly indicates that the IC is not deteriorating signal quality and not introducing significant noise and non-linearity. Measurement results obtained with a 40\,Gb/s, 10\,km SMF link that uses 1550\,nm wavelength carrier show that the equalizer can easily replace the existing ADC+DSP based equalizers for short-reach, high-capacity DCIs. 100\,Gb/s post-layout simulation results of the equalizer IC indicate that its architecture can be used to enhance the capacity of optical links. The same architecture can be used for links with longer channel lengths if the carrier wavelength is 1310\,nm or for links that use dispersion-shifted fibers.

\bibliographystyle{IEEEtran}
\bibliography{./arXiv}

\end{document}